\documentclass[sigconf, nonacm]{acmart}

\usepackage{xcolor}
\usepackage{enumitem}

\usepackage{tabularx}
\usepackage{booktabs}
\usepackage{multirow}
\usepackage{siunitx}
\setlist[itemize]{leftmargin=*}

\usepackage{CJK}
\usepackage{marvosym}

\newcommand\vldbdoi{XX.XX/XXX.XX}
\newcommand\vldbpages{XXX-XXX}
\newcommand\vldbvolume{14}
\newcommand\vldbissue{1}
\newcommand\vldbyear{2020}
\newcommand\vldbauthors{\authors}
\newcommand\vldbtitle{\shorttitle} 
\newcommand\vldbavailabilityurl{https://github.com/elem-azar-unis/CRDT-Redis/tree/master/MET}
\newcommand\vldbpagestyle{plain}

\newcommand{\met}{\textsc{Met}\ }
\newcommand{\mett}{\textsc{Met}}
\newcommand{\rwf}{\textsc{Rwf }}
\newcommand{\rwff}{\textsc{Rwf}}

\newcommand{\send}{\textit{send }}
\newcommand{\rcv}{\textit{receive }}
\newcommand{\doo}{\textit{do }}
\newcommand{\readd}{$re\text{-}add$\ }

\theoremstyle{plain}
\newtheorem{manualissue}{Issue}
\newenvironment{issue}[1]{%
    \manualissue
}{\endmanualissue}

\begin{document}
\title{\textsc{Met}: Model Checking-Driven Explorative Testing of CRDT Designs and Implementations}

%%
%% The "author" command and its associated commands are used to define the authors and their affiliations.

\author{Yuqi Zhang, Yu Huang$^*$, Hengfeng Wei$^*$, Xiaoxing Ma} \thanks{$^*$ Corresponding author.}
\affiliation{%
    \institution{State Key Laboratory for Novel Software Technology, Nanjing University}
    %\streetaddress{P.O. Box 1212}
    %\city{Nanjing}
    %\state{China}
    %\postcode{210023}
    %\institution{$^\text{\textsection}$ Huawei Programming Language Lab}
    %\streetaddress{P.O. Box 1212}
    %\city{Hangzhou}
    %\state{China}
    %\postcode{310052}
}
\email{cs.yqzhang@gmail.com, {yuhuang, hfwei, xxm}@nju.edu.cn}

%\author{Yuqi Zhang}
%\affiliation{%
%    \institution{State Key Laboratory for Novel Software Technology, Nanjing University}
%    %\streetaddress{P.O. Box 1212}
%    \city{Nanjing}
%    \state{China}
%    \postcode{210023}
%}
%\email{cs.yqzhang@gmail.com}
%
%\author{Yu Huang$^*$} \thanks{* Corresponding author.}
%\affiliation{%
%    \institution{State Key Laboratory for Novel Software Technology, Nanjing University}
%    %\streetaddress{P.O. Box 1212}
%    \city{Nanjing}
%    \state{China}
%    \postcode{210023}
%}
%\email{yuhuang@nju.edu.cn}
%
%\author{Xxx Xxx}
%\affiliation{%
%    \institution{Huawei}
%    %\streetaddress{P.O. Box 1212}
%    \city{Hangzhou}
%    \state{China}
%    \postcode{}
%}
%\email{xxx@xxx}
%
%\author{Hengfeng Wei$^*$}
%\affiliation{%
%    \institution{State Key Laboratory for Novel Software Technology, Nanjing University}
%    %\streetaddress{P.O. Box 1212}
%    \city{Nanjing}
%    \state{China}
%    \postcode{210023}
%}
%\email{hfwei@nju.edu.cn}
%
%\author{Xiaoxing Ma}
%\affiliation{%
%    \institution{State Key Laboratory for Novel Software Technology, Nanjing University}
%    %\streetaddress{P.O. Box 1212}
%    \city{Nanjing}
%    \state{China}
%    \postcode{210023}
%}
%\email{xxm@nju.edu.cn}

%%
%% The abstract is a short summary of the work to be presented in the
%% article.

%--
\begin{abstract}

Internet-scale distributed systems often replicate data at multiple geographic locations to provide low latency and high availability, despite node and network failures.
According to the CAP theorem, low latency and high availability can only be achieved at the cost of accepting weak consistency.
The Conflict-free Replicated Data Type (CRDT) is a framework that provides a principled approach to maintaining eventual consistency among data replicas. 
CRDTs have been notoriously difficult to design and implement correctly. 
Subtle deep bugs lie in the complex and tedious handling of all possible cases of conflicting data updates.
We argue that the CRDT design should be formally specified and model-checked, to uncover deep bugs which are beyond human reasoning.
The implementation further needs to be systematically tested. 
On the one hand, the testing needs to inherit the exhaustive nature of the model checking and ensures the coverage of testing.
On the other hand, the testing is expected to find coding errors which cannot be detected by design level verification.

Towards the challenges above, we propose the \underline{M}odel Checking-driven \underline{E}xplorative \underline{T}esting (\mett) framework.
At the design level, \met uses TLA+ to specify and model check CRDT designs.
At the implementation level, \met conducts model checking-driven explorative testing, in the sense that the test cases are automatically generated from the model checking traces.
The system execution is controlled to proceed deterministically, following the model checking trace.
The explorative testing systematically controls and permutes all nondeterministic choices of message reorderings.

We apply \met in our practical development of CRDTs.
The bugs in both designs and implementations of CRDTs are found.
As for bugs which can be found by traditional testing techniques, \met greatly reduces the cost of fixing the bugs.
Moreover, \met can find subtle deep bugs which cannot be found by existing techniques at a reasonable cost. 
Based on our practical use of \mett, we discuss how \met provides us with sufficient confidence in the correctness of our CRDT designs and implementations. 
        
\end{abstract}

\maketitle

%%% do not modify the following VLDB block %%
%%% VLDB block start %%%
\pagestyle{\vldbpagestyle}
\begingroup\small\noindent\raggedright\textbf{PVLDB Reference Format:}\\
\vldbauthors. \vldbtitle. PVLDB, \vldbvolume(\vldbissue): \vldbpages, \vldbyear.\\
\href{https://doi.org/\vldbdoi}{doi:\vldbdoi}
\endgroup
\begingroup
\renewcommand\thefootnote{}\footnote{\noindent
    This work is licensed under the Creative Commons BY-NC-ND 4.0 International License. Visit \url{https://creativecommons.org/licenses/by-nc-nd/4.0/} to view a copy of this license. For any use beyond those covered by this license, obtain permission by emailing \href{mailto:info@vldb.org}{info@vldb.org}. Copyright is held by the owner/author(s). Publication rights licensed to the VLDB Endowment. \\
    \raggedright Proceedings of the VLDB Endowment, Vol. \vldbvolume, No. \vldbissue\ %
    ISSN 2150-8097. \\
    \href{https://doi.org/\vldbdoi}{doi:\vldbdoi} \\
}\addtocounter{footnote}{-1}\endgroup
%%% VLDB block end %%%

%%% do not modify the following VLDB block %%
%%% VLDB block start %%%
\ifdefempty{\vldbavailabilityurl}{}{
    \vspace{.3cm}
    \begingroup\small\noindent\raggedright\textbf{PVLDB Artifact Availability:}\\
    The source code, data, and/or other artifacts have been made available at \url{\vldbavailabilityurl}.
    \endgroup
}
%%% VLDB block end %%%

%-------------------- my text

%--
% \begin{CJK*}{UTF8}{gbsn}

%--
%--
\section{Introduction}

Large-scale distributed systems often resort to replication techniques to achieve fault-tolerance and load distribution \cite{Burckhardt14, Shapiro11a, Enes19}. For a large class of applications, user-perceived latency and overall service availability are widely regarded as the most critical factors. Thus, many distributed systems are designed for low latency and high availability in the first place and resort to \textit{eventual consistency} due to the CAP theorem \cite{Brewer12, Gilbert12}.
Eventual consistency allows replicas of some data type to temporarily diverge, and making sure these replicas will eventually converge to the same state in a deterministic way \cite{Shapiro11a}. The Conflict-free Replicated Data Type (CRDT) framework provides a principled approach to maintaining eventual consistency \cite{Shapiro11a, Burckhardt14}. 
CRDTs are key components in modern geo-replicated systems, such as Riak \cite{Riak}, Redis-Enterprise \cite{Redis-Enterprise}, and Cosmos DB \cite{CosmosDB}.

It has been notoriously difficult to correctly design and implement CRDTs.
CRDT implementations suffer from so-called \textit{deep bugs} \cite{Lee14}.
From the extensional perspective, deep bugs often appear in rare situations, but they still manifest themselves, often in critical situations, in large or longtime deployments.
From the intensional perspective, due to the intrinsic uncertainty in distributed system execution, deep bugs only appear when the system is faced with certain subtle combinations of system and environment events.
Though the bug triggering patterns of events usually involve only a moderate number of events, the space of all potential bug triggering patterns is exponentially large.
This explains why deep bugs are hard to replay and why they are often beyond the coverage of standard testing techniques. 

%\mybreak

It is a great challenge to prevent, detect and fix deep bugs in the design of a CRDT.
The central issue in CRDT design is to resolve conflicts between concurrent updates, no matter how the updates are out of order on different replicas.
Design of the conflict resolution strategy is typically tedious and error-prone, especially for data types with complex semantics. 
The designer has to exhaustively check all possible interleavings of conflicting updates, which results in a great amount of metadata.
The metadata needs continuous and consistent maintenance.
Existing CRDT designs are mainly informal. 
The correctness of the design highly depends on the experience of the designer. The design also needs intensive design review.
All these factors make the design process quite time- and energy-consuming.

In order to cope with subtle deep bugs in the design of a CRDT, it is necessary to have a precise and unambiguous description of the design. 
More importantly, the design should be "debuggable". 
That is, the design needs to be automatically explored by a machine.
All corner cases can be covered by the exhaustive exploration.
Moreover, when a bug is found in the design, the exploration process is also expected to provide sufficient information for the designer to find the root cause of the bug and then fix it.
%This process is repeated until the design is sufficiently correct.

It is also a great challenge to cope with deep bugs in the implementation of a CRDT, even after the design is deemed correct by formal verification.
%As for the implementation of CRDTs, we still have severe problems even the design is deemed correct.
First, the implementation has to handle details which are not covered in the design. 
The transcription from the design to the implementation often introduces subtle bugs.
Moreover, the design is intended for human reasoning. The developer may often find the design inefficient when directly translated to the implementation.
Thus the developer often has a strong incentive to optimize the design. Such intentional deviations often greatly increase the odds of transcription errors in the implementation.

Second, the design often makes "reasonable" assumptions.
In CRDT implementations, such assumptions have to be carefully ensured by extra implementations which are neglected in the design.
The implementations centering around assumptions in the design are expected to be straightforward and highly dependable.
However, this is often not the case when implementing complex CRDTs.
For example, ensuring an assumption in the design often makes use of third-party libraries. Such libraries often have many options for different usage in different scenarios.
Misconfiguration or misuse of the libraries may introduce subtle bugs.
Such bugs often evade human reasoning and code review since the corresponding implementations are not specified in detail in the design.

Existing testing techniques are insufficient to cope with deep bugs in CRDT implementations. 
Code-level model checking for distributed systems can handle deep bugs, but usually imposes a prohibitive cost.
What we need is to achieve the best of both worlds. We need the ability of distributed system model checking to exhaustively check all corner cases of system execution, but also need the cost-effectiveness of standard testing techniques.

Toward the challenges above, we propose the \underline{M}odel Checking-driven \underline{E}xplorative \underline{T}esting (\mett) framework. %to achieve the best of both worlds.
As indicated by its name, the \met framework consists of two layers.
In the \textit{design layer} (also denoted the \textit{model layer}), \met employs the Temporal Logic of Actions (TLA+) for the specification of CRDT protocols.
%we specify the CRDT design in TLA+ (Tempoal Logic of Actions). 
TLA+ is a lightweight formal specification language especially suitable for the design of distributed and concurrent systems \cite{TLA}. %Leveraging simple math, TLA+ can express concepts much more elegantly and accurately.
The specification in TLA+ precisely describes what is required (i.e., specification of the correctness properties) and what is to be implemented (i.e., specification of the CRDT protocol).
More importantly, the design specified in TLA+ is model-checked to eliminate any violation of the user-specified correctness conditions.
The key to practical and effective model checking is to tame the state explosion problem. 
\met coarsens the specification of the CRDT protocol to prune unnecessary interleavings of events.
\met also directly limits the scale of the system model to reduce the model checking cost.
\met leverages the designer's understanding of the CRDT protocol to ensure that the pruning of the state space will not significantly hamper the ability of \met to "dig out" deep bugs.

%\met provides principles for making tradeoffs between the confidence gain by the model checking and the checking cost required.
%certain details which are not relevant to the deep bugs are omitted.
%The scale of the CRDT store is limited.

%Error trace can be used to find the root cause and fix the bug.

In the \textit{implementation layer} (also denoted the \textit{code layer}), \met conducts \textit{explorative testing} on CRDT implementations, in the sense that the testing systematically controls and permutes all nondeterministic choices of message reorderings.
%\met automatically generates test cases guided by the model checking in the design layer, thus 
\met inherits the exhaustive exploration from the design layer to the implementation layer.
The explorative testing contains two key components: generation of test cases and enhancement of system testability.

%1.1
As for test case generation, the test cases are automatically extracted from the model checking trace. 
The model checking trace consists of a sequence of system states connected by events triggering the state transitions.
The sequence of events, denoted as the event schedule, is extracted as the test input. The events mainly include system events, e.g., client requests, synchronization among replicas, and environment events, e.g., (out-of-order) message delivery.
%As for the CRDT we focus on, the uncertainty is message reorderings.
% 1.2
The sequence of system states is extracted as the test oracle. That is, the code-level execution is expected to produce the same sequence of system states as the model-level execution.

% 2.1
As for enhancement of system testability, \met first enhances system controllability. That is, \met deterministically replays the model checking trace in the implementation layer. This is achieved by hacking the underlying RPC among server replicas of the CRDT store.
The communications among server replicas are all directed to a central test manager. 
The test manager discharges the intercepted events one by one, strictly following the event schedule in the test input.
%decides the ordering of all events intercepted according to the model checking trace.
Note that this hacking is transparent to the upper layer CRDT implementation.

% 2.2
Besides system controllability, \met further needs to enhance system observability. 
\met does not adopt the typical approach, where the system states are logged and then analyzed offline after the test case execution.
\met first implements dedicated APIs for the test manager to inspect the internal states of the server replicas.
Given the controllability of the system, the test manager intercepts all system events and decides the total order of the events.
Then the test manager inserts the system state inspection commands into the event schedule to record the system state.
This approach greatly simplifies the work of comparing the state sequence of system execution with that of the model checking, i.e., the test oracle.

%\met provides the metrics derived from the oracle. 
%\met augment the system with APIs for explorative testing. The tester directly query state via the API at runtime.

%also provides how to instrument the system with logging.

%In contrast to the standard testing methods, explorative testing need to control the execution. all uncertainties are to be determined. This is instructed by the model checking.
%To execute the test case, we need to enhance the testability of the system.

%we model checking traces to generate test cases.
%The test cases systematically control and permute non-deterministic choices such as message reorderings.
%The test case find code level bugs, such as bugs pertaining to framework and implementation details not covered in the design.
%
%Then the conformance of the code level to the model level shows the correctness of the implementation, with respect to the finite model and our simplifying assumptions.
 
%\mybreak
 
%【实现和实验】

The \met framework is applied in the design and implementation of different types of Replicated Priority Queues (RPQs) and Replicated Lists (RLists) over the CRDT-Redis data type store \cite{CRDT-Redis}.
We first discuss how our experiences in testing CRDT implementations motivate the \met framework.
Second, we discuss the bugs found by \met both in both design and implementation of CRDTs.
The \met framework greatly eases the fixing of bugs which can be detected by standard testing techniques. 
\met also finds bugs which cannot be detected by standard testing techniques.
Third, we discuss how the exhaustive (constrained by the testing budget) and explorative testing of CRDTs using \met increases our confidence in the correctness of the design and implementation of CRDTs.

%\sucai{In the MET framework, we first in the model level use TLA+ to specify the design. The design is then model checked to prevent bugs. Then, we use the exhaustively checked trace in the model level as the correctness specification for the underlying CRDT implementations. We control the execution to exhaustively execute the traces in the code level. In order to tackle the state explosion problem, we introduce reasonable assumptions and omit non-essential details. With respect to our assumption, we exhaustively checks the code and shows its correctness. }

%We apply MET on the Replicated Priority Queue (RPQ) and Replicated List (RL) implementations. MET model checks the design. MET exhaustively checks all possible executions of real codes. We derive reasonable assumptions from human experience and domain knowledge. The assumptions help to reduce the checking cost. The application of MET on these two concrete examples shows the effectiveness of MET.

The rest of this work is organized as follows. 
Section \ref{Sec: Moti-Ovw} extensively discusses the motivation for \met and overviews of the design of \mett. 
Section \ref{Sec: Model-Level} and \ref{Sec: Code-Level} present the model layer and the code layer design of \met respectively. Section \ref{Sec: Exp} demonstrates the application of \met in practical CRDT design and implementation. Section \ref{Sec: RW} discusses the related work. In Section \ref{Sec: Concl}, we conclude this work and discuss the future work.

\section{Motivation and Overview} \label{Sec: Moti-Ovw}

In this section, we first present our practical experiences in CRDT development, in order to extensively explain the motivation behind \mett.
Then we provide an overview of \mett.

%--------------------------------------------------------------------------------
\subsection{Our Experiences Motivating \mett} \label{SubSec: Expe-Moti}

The \met framework is proposed in our practices in developing different types of CRDTs.
Specifically, we propose the \rwf conflict resolution strategy to ease the design and implementation of data container CRDTs.
\rwff-RPQ and \rwff-List are developed on the CRDT-Redis data type store \cite{CRDT-Redis}.
For the purpose of performance comparison, we also develop the RemoveWin-RPQ and RemoveWin-List.
\rwf and RemoveWin can be viewed as two different types of conflict resolution strategies.
The details of these strategies are irrelevant to our discussions on the \met framework here.
More details of these CRDTs can be found in \cite{Zhang21}.

The key challenge in CRDT development is to ensure that the conflict resolution strategy guarantees eventual consistency, no matter how the data updates are out of order.
Lacking an explorative testing service or tool, we resort to random stress testing.
We keep generating concurrent and conflicting data updates and randomly dispatch the updates to all the replicas.
We further use traffic control (TC) \cite{TC20} to add random delay to the messages among server replicas to make the updates out of order.

The testing proceeds in rounds.
In each round, 10,000 update operations are generated per second, lasting 5 minutes.
We check whether the replicas reach the same state after they have received all the data updates at the end of each round.
A bug is found if any two replicas do not reach the same state.
If no bug is found after 24 hours of stress testing, the implementation is deemed (sufficiently) correct.
More details of the CRDT development can be found in our previous work \cite{Zhang21}.
The experiences in testing our CRDT implementations extensively motivate us to design the \met framework, as detailed below.

%----------------------------------------
\subsubsection{Why We Need Debuggable CRDT Design} \label{SubSubSec: Why-MCK}

The conflict resolution logic in a CRDT protocol is usually quite complex and error-prone.
The informal design of a CRDT protocol is far from being enough to ensure the correctness of the design.
Informal software designs, e.g., textual descriptions, flow charts and pseudo-codes, are dominant in current software development practices.
The correctness of informal software design is mainly guaranteed by manual reasoning and intensive design review among a team of developers.
However, it is widely accepted that human intuition is poor at estimating the true probability of supposedly extremely rare combinations of events in real deployments of complex distributed systems \cite{Lee14, Newcombe15}.

For example when developing the \rwff-List \cite{Zhang21}, we find violations of eventual consistency through stress testing in a small number of experiments, as shown in Issue \ref{I: Iss1} below.

%--
\begin{issue}{1}\label{I: Iss1}
    List replicas may not converge in the stress testing \rm{\cite{Issue1}}. 
\end{issue}

\noindent To find the root cause of this bug, we first look at the final states of two replicas, and identify the list elements with divergent positions.
As specified in the \rwff-List protocol, the position of one element is decided when the element is added to the list for the first time.
Thus we go backtracking through the execution trace, in order to find the point when the list replicas first diverge, and the point when the elements with divergent positions are first added to the list.
In this way, we find that the bug is caused by the element which is in the list but has an undefined position.

\begin{figure}[htbp]
    \centering
    \includegraphics[width=\linewidth]{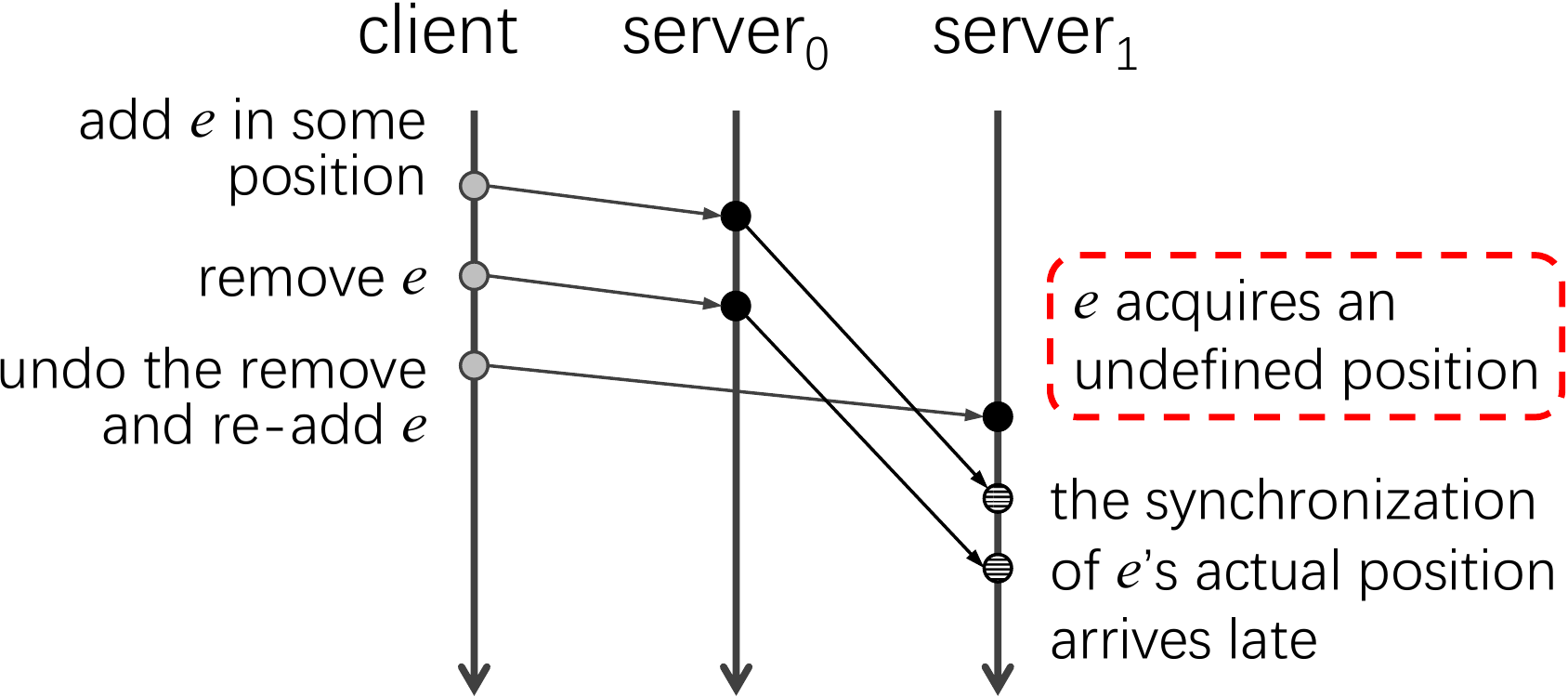}
    \caption{Buggy design of the re-add operation.}
    \label{F: Undo-Redo-Existence}
\end{figure}

A minimized example of this bug is shown in Figure \ref{F: Undo-Redo-Existence}.
According to our design of the \rwff-List, the location of list element $e$ is decided when it is first added on $server_0$. 
If we remove and then re-add $e$ (such as undoing the removal of words in collaborative editing), its position will not change. 
The client may send the $re\text{-}add(e)$ operation to $server_1$, which has not yet gotten the previous $add(e)$ operation from $server_0$. 
$Server_1$ should reject this \readd operation or buffer this operation for processing later.
But in our buggy design, $server_1$ accepts this request and assigns an illegal position to $e$. 
When $server_1$ receives the delayed $add(e)$ operation, it will ignore this operation, which causes the divergence among replicas.

%--

Though the debugging process above is methodologically simple, it is quite time- and energy-consuming.
The stress testing runs for hours or even days. 
The error trace is quite long when the divergence among replicas is identified.
Both trace analysis scrips and manual inspection are needed to find when the divergence first occurred and when the elements corresponding to the divergence were first added to the list.
Given the information about the divergence, we then need to manually reason how the elements are handled by the replicas, in order to understand how the bug is introduced in the design and the implementation.
We spent several days to find the root cause and fix this bug.

Note that in our design, the \readd operation is treated as a special case of the \textit{add} operation.
To fix the bug in Issue \ref{I: Iss1}, we add one more case in the conflict resolution logic of the \textit{add} operation.
The resulted implementation then passes the stress testing.
However, we highly suspect that there are still deep bugs hidden in our design.
It is mainly because the conflict resolution logic in the \textit{add} operation is quite complex. 
Further adding more cases in the already-quite-complex logic in \textit{add} makes the design highly unreliable.
Manual reasoning is powerless to ensure the correctness of this complex design.

The experiences above convince us to add formal specification and verification as one obligatory step in our CRDT design.
Though formal specification also requires non-trivial human efforts, the specification process will force the designer to be unambiguous about the details in the design. The specification can further be exhaustively checked by a model checker.
We expect the increase in design quality and decrease in testing cost can well compensate for the human efforts in the formal specification process.

%----------------------------------------
\subsubsection{Why We Need Code-level Testing} \label{SubSubSec: Why-Testing}

Even though the design has been deemed correct after the formal verification, we still need code-level testing.
It is mainly because, at the model level, we verify a system model, not the actual system implementation.
The obtained results are thus as good as the system model.
Different types of coding errors can be introduced in the implementation.
We mainly discuss two salient types of coding errors, namely the \textit{transcription error} and the \textit{assumption error}.

We first discuss the transcription errors, i.e., errors introduced when transcribing the design into the code.
As we know, software design is intentionally abstract. The developer must instantiate the abstract design and transform it to correct executable code.
The developer will meet numerous details which are not covered in the design.
This process is often error-prone.
Moreover, the developer often has the incentive to optimize the design in the implementation. 
This is mainly because that the informal design is intended for human reasoning. Thus directly translating the design to executable code is often a feasible but inefficient choice. 
Even for the formal design intended for machine exploration, there is still a significant gap in semantics between a formal specification and an executable program.
Thus the programmer may often try to find an equivalent but more efficient implementation from the abstract design.
This process often increases the odds of introducing transcription errors.

Though our random stress testing did not find bugs due to transcription errors (denoted the \textit{transcription bugs}), we highly suspect that there must be such bugs in our implementation. We "borrow" a transcription bug found by the \met framework from Section \ref{SubSubSec: Trans-Bug} for the purpose of illustration here.

%--
\begin{issue}{4} \label{I: Iss4}
    In CRDTs developed using the \rwf framework, the existence of one element is protected by a precondition in the design. However, the implementation deviates from the design in the case when the precondition does not hold \rm{\cite{Issue4}}.
\end{issue}

\begin{figure}[htbp]
    \centering
    \includegraphics[width=\linewidth]{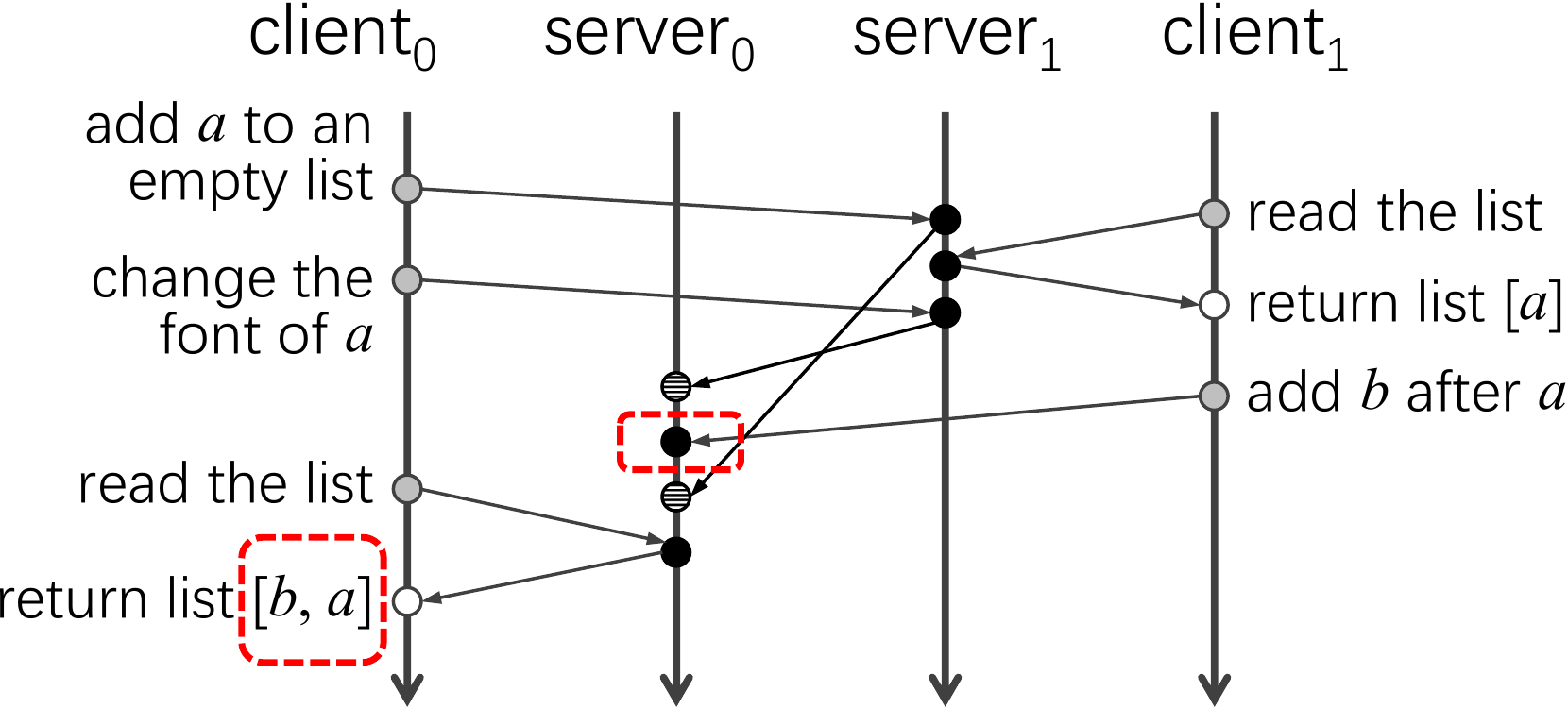}
    \caption{An exemplar transcription bug.}
    \label{F: Trans-Bug}
\end{figure}

\noindent Issue \ref{I: Iss4} can be illustrated by the example in Figure \ref{F: Trans-Bug}.
In this example, $client_0$ first adds element "a" and then changes the value of "a" (say, it changes the font of character "a" in a string in a collaborative editing scenario).
Then $client_1$ queries the value of the string and obtains string "[a]".
After reading the value of the string, $client_1$ inserts "b" after "a" into the string.
However, in our buggy implementation, $client_0$ may read string "[b,a]".

This bug can be triggered by the following adversarial pattern of events.
%The operation "adding $a$ to an empty string" is intentionally delayed for a sufficiently long period of time on $server_0$ (not drawn in the figure).
$Server_0$ first receives the operation "changing the font of $a$" (without adding $a$ first). 
Then $server_0$ will receive the operation "adding $b$ after $a$".
At the design level, this "adding $b$ after $a$" operation is quite safe, since it is protected by the precondition that "before adding $b$ after $a$, $a$ must exist".
The design does not explicitly state what to do if the precondition does not hold.
The precondition implicitly states that, if the element does not exist, nothing should be done.
However, in our implementation, an illegal position is assigned to $b$, and then to $a$.
In this way, $client_0$ gets "$[b,a]$".
This bug is beyond the coverage of model checking in the design layer, and we discuss in Section \ref{SubSubSec: Trans-Bug} how we detect this bug with the help of \mett.

%--

The second type of coding errors, namely the assumption error, pertains to assumptions in the design.
In a broader sense, the assumption error is also a kind of transcription error.
However, as various assumptions are widely used in designs of distributed protocols and systems, we explicitly separate this type of errors out.

In designs of distributed protocols and systems, many details are often intentionally omitted, via the form of "reasonable" assumptions.
The assumptions are usually assumed to be straightforward to implement or to be readily provided by existing libraries/tools. 
Thus details of how the assumptions should be guaranteed are omitted in the design.
However, when different modules of a system have different or even conflicting assumptions, the developer needs to put more effort into managing the implementations pertaining to a number of various assumptions.
Moreover, third-party libraries or tools can have complex semantics and subtle configurations. 
Even if the library implementation is perfectly correct, the developer may still use it with a wrong configuration, thus failing to provide the intended guarantee required by the assumption.
In such situations, the developers may introduce assumption errors unwittingly.
As the number of assumptions in the design increases, the odds of \textit{assumption bugs}, i.e., bugs due to the assumption errors, quickly increase to an extent that cannot be neglected.

\begin{figure}[htbp]
    \centering
    \includegraphics[width=0.85\linewidth]{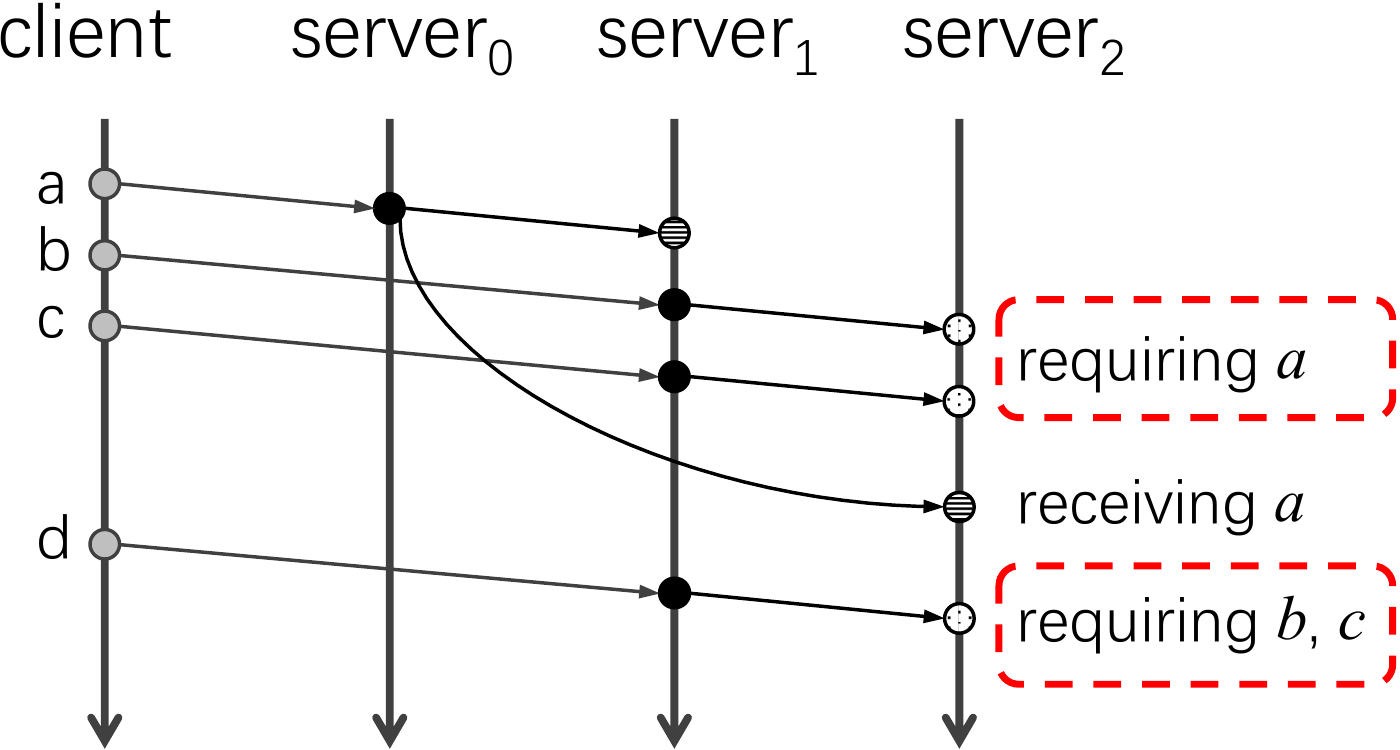}
    \caption{An exemplar assumption bug.}
    \label{F: Assumption-Bug}
\end{figure}

Issue \ref{I: Iss2} is an example of the assumption bug in our implementation.
The RemoveWin CRDTs assume that the underlying network provides the causal delivery semantics,
but the networking primitives of the CRDT-Redis data store do not provide such semantics.
In contrast, the \rwf CRDTs do not need the causal delivery network. 
In the example shown in Figure \ref{F: Assumption-Bug}, the $client$ sends requests $a$, $b$ and $c$ to different servers.
$Server_1$ receives $b$ and $c$ directly from the $client$, and receives the synchronization of $a$ from $server_0$.
Since the synchronization of $a$ arrives at $server_1$ before that of $b$ and $c$, $server_1$ works correctly. 
In contrast, $server_2$ gets $b$ and $c$ from $server_1$ first, without getting $a$.
The upper layer RemoveWin CRDT protocol assumes that the underlying network provides causal delivery of messages.
Thus the CRDT protocol does not consider the case in our example and $b$ and $c$ are erroneously processed on $server_2$.

The root cause behind this bug is that the \rwf CRDTs and the RemoveWin CRDTs have different assumptions on the underlying network.
The CRDT-Redis platform aims to host different CRDT protocols, which may have different assumptions on the network, the persistent storage, etc.
Thus we need to "align" the assumptions of different protocols on the CRDT-Redis platform.
This situation is similar to the misconfiguration problem in cloud and data center platforms \cite{Xu16}.
The implementations pertaining to such alignments are error-prone and are not sufficiently documented in the design.

Also note that, even when we find the root cause of an assumption bug, the fixing of this type of bugs is often non-trivial.
In our first patch to fix Issue \ref{I: Iss2}, $server_2$ does not synchronize $b$ and $c$ correctly when it receives $a$.
Then when $server_2$ later receives $d$, it gets stuck since the required delivery of $b$ and $c$ are not available.
We argue that not only the original implementation, but also the patches fixing the assumption bugs need extensive testing.

%--
\begin{issue}{2}\label{I: Iss2}
    The RemoveWin CRDT protocols assume that the underlying network provides causal delivery of messages.
    The CRDT-Redis platform does not provide such semantics of network communication \rm{\cite{Issue2}}. 
\end{issue}

%--

Besides the underlying network, the CRDT protocols may also have assumptions about other modules in the system.
More assumptions bugs will be discussed in Section \ref{SubSubSec: Assum-Bugs}.

%----------------------------------------
\subsubsection{Why We Need Automatic Explorative Testing}

The Issue \ref{I: Iss2} discussed above is discovered by manual explorative testing.
Given the effectiveness of this manual explorative testing, we decide to automate this process.
This automation gives us the \met framework.
Specifically, we, as the designer of the CRDT protocol, know which part of the conflict resolution is most tricky and unreliable.
Thus we manually construct adversarial executions which target at the most tricky part of our design.
Given the adversarial executions, we then extend our unit testing to replay such executions for one server replica.
The inputs are given to the replica following the adversarial execution.
The replica then executes its CRDT protocol.
When the replica interacts with the outside world, we manually calculate what should be returned to the replica.
The server replica is provided with the illusion that it obtains the feedback from other peer replicas. %, while actually this results are from our manual calculation and reasoning.

Issue \ref{I: Iss2} is out of reach of stress testing and is found by this manual explorative testing.
%
%Though the manual explorative testing process is quite inefficient, it exercises the CRDT implementation with the most important test cases, having the highest probability to find a deep bug if any.
%We actually find bugs which cannot be found by stress testing, which is Issue \ref{I: Iss2}.
%
We can easily see the advantages and disadvantages of the manual explorative testing. 
The main advantage is that, the test cases are significantly more effective than the randomly generated ones.
The main disadvantage is that, the whole process is manual and imposes a prohibitive cost. 
Our CRDT implementations cannot be sufficiently tested using this manual testing.

The \met framework is directly shaped by this manual explorative testing practice.
\met enumerates all possible test cases, thus deterministically covering the most important ones.
\met automates the most tedious part of the manual testing and enables efficient explorative testing.
%We overview \met below.

%include, the test case is of special quality. It is good at finding deep bugs.
%The testing is not random. It follows the good test case.
%%
%The disadvantages include, this process is manual and quite depend on the user experience.
%The construction of the adversarial execution is quite non-trivial.
%The manual reasoning and calculation of what should be returned to the replica under test is not hard, but is quite energy consuming.

%--------------------------------------------------------------------------------
\subsection{Overview of the \met Framework} %\label{}

\begin{figure}[htbp]
    \centering
    \includegraphics[width=0.95\linewidth]{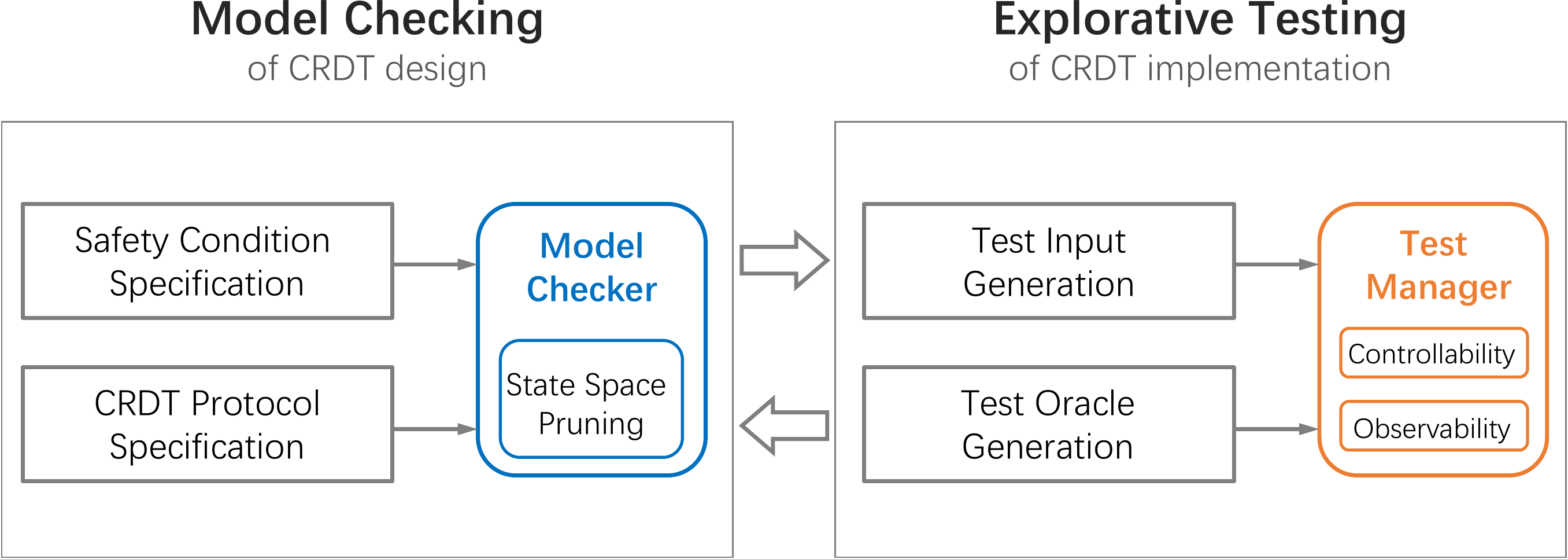}
    \caption{Architecture of the MET framework.}
    \label{F: Framework}
\end{figure}

The \met framework consists of two layers.
The architecture of \met is illustrated in Figure \ref{F: Framework}. 

%--

\noindent In the model layer, we have:
%--
\begin{itemize}
    \item (Section \ref{SubSec: TLA-Spec}) \textit{Specification}. The CRDT protocol and the correctness conditions are formally specified in TLA+.
    \item (Section \ref{SubSec: MCK}) \textit{Model Checking}. The specifications are automatically explored by the TLC model checker. 
        %When violations are detected by model checking, the error traces are provided, which facilitate the fixing of the bug. 
        Reduction techniques are proposed to tame the state explosion problem.
\end{itemize}

\noindent In the code layer, we have: %model checking-driven explorative testing. The testing has two essential parts:
\begin{itemize}
    \item (Section \ref{SubSec: Test-Case-Gen}) \textit{Test Case Generation}. Test cases are automatically generated from the model checking trace.
        Test case generation includes generation of the \textit{test inputs} and that of the \textit{test oracle}. %two parts: test input generation and test oracle generation.
    \item (Section \ref{SubSec: Testability}) \textit{Testability Enhancement}. 
        %The system execution is under full control of the test manager. 
        The test manager replays the model checking trace in the code layer, which requires the enhancement of system \textit{controllability} and \textit{observability}.
        %To enable this explorative testing, the system under test should be enhanced with sufficient testability, including controllability and observability.
\end{itemize}

\section{Formal Specification and Model Checking of CRDT Designs} \label{Sec: Model-Level}

In this section we introduce the formal verification of CRDT design in \mett, including formal modeling in TLA+ and taming of the state explosion problem in model checking.

%--------------------------------------------------------------------------------
\subsection{Formal Modeling in TLA+} \label{SubSec: TLA-Spec}

The formal modeling includes the specification of the CRDT protocol and that of the correctness properties. The formal specification language we use is TLA+.

%----------------------------------------
\subsubsection{TLA+ Basics}

In TLA+ we model a distributed system in terms of one single global state. A distributed system is specified as a state machine by describing the possible initial states and the allowed state transitions called $Next$. 
The system specification contains a set of (global) system variables $V$. A \textit{state} is an assignment to the system variables. $Next$ is the disjunction of a set of actions $a_1\lor a_2\lor \cdots \lor a_p$, where an \textit{action} is a conjunction of several clauses $c_1\wedge c_2\wedge \cdots \wedge c_q$.  A \textit{clause} is either an \textit{enabling condition}, or a \textit{next-state update}. 

An enabling condition is a state predicate which describes the constraints the current state must satisfy, while the next-state update describes how variables can change in a step (by "step" we mean successive states).
Whenever every enabling condition $\phi_a$ of an action $a$ is satisfied in a given "current" state, the system can transfer to the "next" state by executing $a$, assigning to each variable the value specified by $a$. 
We use "$s_1\stackrel{a}{\rightarrow}s_2$" to denote that the system state goes from $s_1$ to $s_2$ by executing action $a$, and $a$ can be omitted if it is obvious from the context.
Such execution keeps going and the sequence of system states forms a trace of system execution. %All such traces are all possible system behaviors. 
Exemplar TLA+ specifications can be found in the following Figure \ref{F: TLA-RPQ-Spec} and Figure \ref{F: TLA-List-Invariant}. 

One salient feature of TLA+ is that correctness properties and system designs are just steps on a ladder of abstraction, with correctness properties occupying higher levels, system designs and algorithms in the middle, and executable code and hardware at the lower levels \cite{Newcombe15}. This ladder of abstraction helps designers manage the complexity of real-world distributed systems. Designers may choose to describe the system at several "middle" levels of abstraction, with each lower level serving a different purpose, such as to understand the consequences of finer-grain concurrency or more detailed behavior of a communication medium. The designer can then verify that each level is correct with respect to a higher level. The freedom to choose and adjust levels of abstraction makes TLA+ extremely flexible. 

%For example, a low-level specification for leader election mechanism of Raft may accurately describe how an eligible server is selected as leader through voting, while a high-level one may directly assign some eligible server to be leader and leave the details of voting unspecified.

%----------------------------------------
\subsubsection{Specification of CRDT Protocols} \label{SubSubSec: Proto-Spec}

A replicated data type store consists of a group of replicas and clients, as well as the network channel that connects them.
The state of the replica consists of: i) the concrete data of the data type, ii) the meta-data for conflict resolution, and iii) the input / output message buffer for network communication. 
The essential issue of specifying a CRDT protocol is to specify how the replica updates its state.
The state transfer can be driven by three kinds of events, as defined in the generic framework of CRDT design \cite{Shapiro11a, Shapiro11b}:
%--
\begin{itemize}
    \item \send: the \send event can be triggered when the output buffer of the server replica is not empty. It will broadcast one message in the output buffer to all other replicas.
    
    \item \rcv: the \rcv event can be triggered when there are messages to be delivered from the network channel. It will deliver one message from the channel to the input buffer of the server replica.
    
    \item \doo: the \doo event can be triggered when the input buffer is not empty. It consumes one message from the input buffer and updates the state of the server replica. If the message is a client request, the \doo event will further put a synchronization message into the output buffer of the replica.
\end{itemize}

%\noindent - \textit{send}: the \textit{send} event can be triggered when the output buffer is not empty. It will broadcast one message in the output buffer to all other replicas.
%    
%\noindent - \textit{receive}: the \textit{receive} event can be triggered when there are messages to be delivered from the network channel. It will deliver one message from the channel to the input buffer.
%
%\noindent - \textit{do}: the \textsf{do} event can be triggered when the input buffer is not empty. It consumes one message from the input buffer and updates the state of the replica. If the message is a client request, it will further put a synchronization message into its output buffer.

\noindent Detailed design of the replica update logic is specified in the CRDT protocol. The event-driven design of a CRDT protocol can be readily transformed to TLA+ specifications. More examples of CRDT protocol specifications can be found in our online repository \cite{CRDT-Redis}. %\footnote{https://github.com/elem-azar-unis/CRDT-Redis/tree/master/MET/TLA}.

The underlying network is modeled as a shared global channel, described by a global variable in TLA+. The channel buffers all the messages in transmission.
We assume that the network channel provides the eventual-delivery-once semantics. 
The messages can be out-of-order, but cannot be dropped, duplicated or forged. 
%Eventually all messages will be successfully delivered exactly once.
We also assume that the replicas will not crash.
Note that this assumption is indicated in the definition of eventual consistency \cite{Shapiro11a} and is widely adopted in CRDT designs.
Put it in another way, when the replicas can crash or the messages can be dropped, the data type store does not need to guarantee anything and still provides eventual consistency. Thus such cases are usually not covered in the design of a CRDT protocol.

\begin{figure}[htbp]
    \centering
    \includegraphics[width=\linewidth]{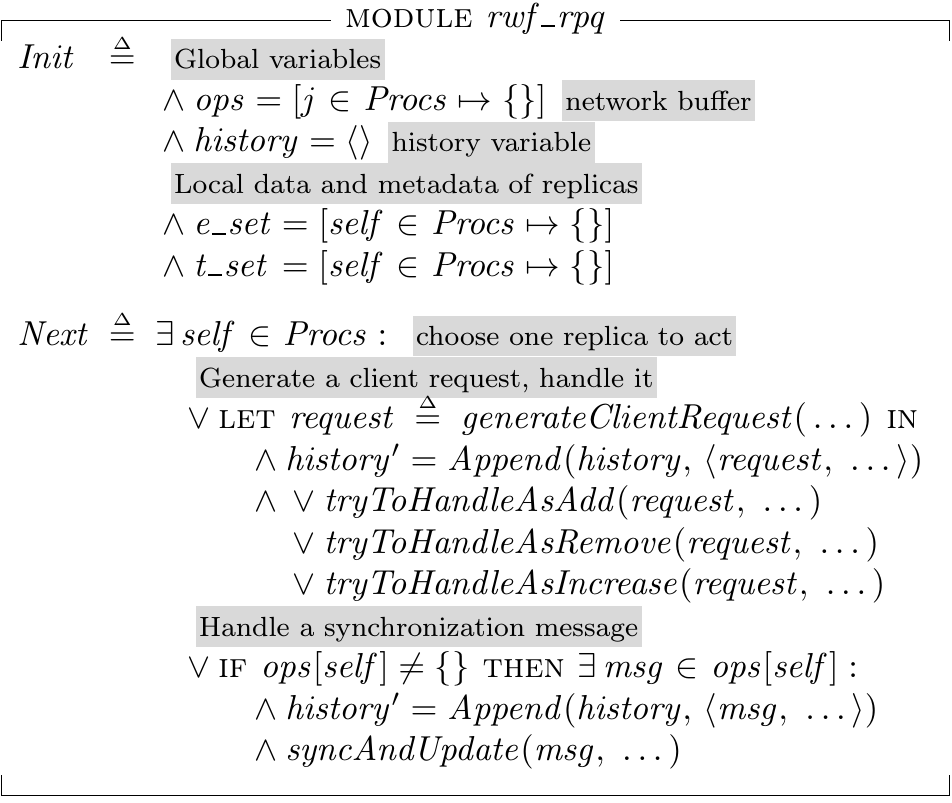}
    \caption{A skeletal example of the TLA+ specification for \rwff-RPQ.}
    \label{F: TLA-RPQ-Spec}
\end{figure}

Given the modeling of the network channel, we do not explicitly model the clients. All client requests are directly modeled as data access messages in the network channel, which will be delivered to the replicas some time in the future. %, as specified by the workload generation logic.
Note that the specification of system behavior intentionally omits certain details. We will justify these simplifications in Section \ref{SubSec: MCK}.
Figure \ref{F: TLA-RPQ-Spec} is a skeletal example of the specification of a CRDT protocol. 
The global variables record status of the server replica. 
The \textit{history} variable records the state transitions (the history variable is introduced for the purpose of test case generation, as detailed in Section \ref{SubSubSec: Gen-Input}). 
The \textit{action} records all possible state transitions.
The model checker simply applies all possible state transitions and enumerates all reachable system states.

%The specification consists of the global variables and the actions.
%Using variables, we record concrete state of the CRDT, metadata for conflict resolution and state of the network channel.
%The action basically states that, every replica is possibly selected to go one step forward.
%The \textit{Next} action considers a number of cases. 
%In each case, the replica executes each clause if its enabling condition is true.

%--

%----------------------------------------
\subsubsection{Specification of Correctness Properties}

The correctness condition we adopt for CRDT design is eventual consistency\footnote{The term eventual consistency may have slightly different meanings in different contexts. In this work, by eventual consistency we mean strong eventual consistency defined in \cite{Shapiro11a}.}.
A CRDT protocol guarantees eventual consistency if any two replicas always converge to the same state as long as they have received the same set of updates, regardless of the order of updates received. The convergence among replicas means that they reach the same state in regard to both the data type itself and the metadata for conflict resolution.

\begin{figure}[htbp]
    \centering
    \includegraphics[width=\linewidth]{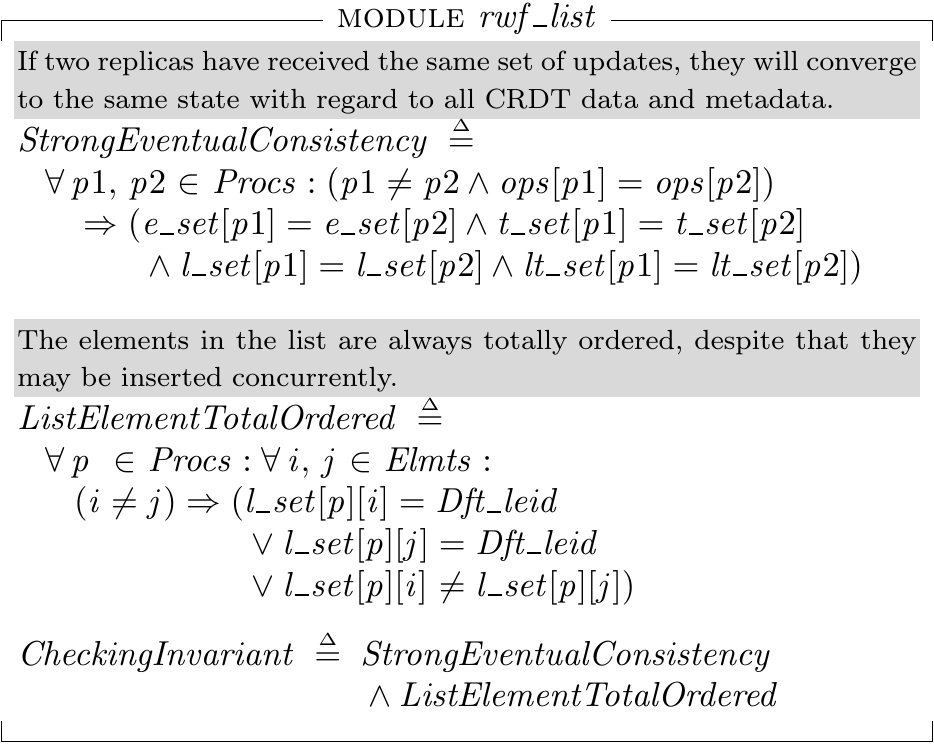}
    \caption{Correctness condition of \rwff-List in TLA+.}
    \label{F: TLA-List-Invariant}
\end{figure}

Besides eventual consistency, we may further specify correctness conditions pertaining to the semantics of the data type under concern.
For example in the design of the \rwff-List, we check whether the position identifiers of all elements in the string are unique and totally ordered.
Figure \ref{F: TLA-List-Invariant} is an example of specifying the correctness conditions for \rwff-List. 
The correctness condition includes both specification of eventual consistency and that of the constraints on the position identifiers.

%--

%There are two aspects of the invariant. 
%First is the strong eventual consistency that two replicas will converge if they have received the same set of updates. By converge we mean that all data and metadata should be the same.
%Then the invariant that the element position identifier in the list are always totally ordered and unique. This means that all the elements in the RWF-List are always totally ordered.

%--------------------------------------------------------------------------------
\subsection{Model Checking of CRDT Designs} \label{SubSec: MCK}

Given the CRDT protocol specified in TLA+, we can now explore the design using the TLC model checker.
Though the model checking process is fully automatic, it is intrinsically constrained by the state explosion problem. Full checking of the system specification of arbitrary scale is usually impractical.

In order to conduct practical model checking of the design and obtain sufficient confidence in the correctness of the design, we prune the state space in advance, from two orthogonal dimensions.
First, the specification is coarsened to omit unnecessary details. 
Second, we leverage the small scope hypothesis to check a small scale system while not significantly sacrificing the effectiveness of finding deep bugs.
%These two types of pruning are discussed in detail below.

%----------------------------------------
\subsubsection{Coarsening the Specification} \label{SubSubSec: Coar-Spec}

We utilize the feature that TLA+ enables the developer to flexibly adjust the level of abstraction.
In the TLA+ specification of a CRDT protocol, the replica mainly processes three types of events. The \send and \rcv events model the network communication. The \textit{do} event mainly updates the metadata for conflict replication and updates the data type itself.
In light of the state explosion problem, we coarsen the TLA+ specification and focus on the \textit{do} event.

We bind the \textit{send} event after the \textit{do} event. 
This means that we let a replica broadcast the synchronization message to all other peer replicas right after it has handled a data update from some client, rather than handle other events first and broadcast the data update in the future. 
Similarly, we bind the \rcv event before the \doo event. This means that we let a replica handle the update request in a message right after the message is received. %, rather than process the update in the future.

This coarsening is mainly based on the observation that the subtle and deep bugs in CRDT designs and implementations mainly result from the complex (and often tedious) processing of diverse patterns of conflicting data updates. 
When we "glue together" the \rcv, \doo and \send events, unnecessary interleavings of events are pruned from the state space, while all possible cases of conflicting resolution are preserved.
As further evidenced by our experimental evaluation on \mett, our coarsening of the specification effectively limits the cost for model checking, while not significantly sacrificing the ability to find deep bugs.

%----------------------------------------
\subsubsection{Limiting the Scale of the Model}

In theory, we need to set the scale of the model to a number larger than any possible instance of the system deployed in a real scenario.
Then the model checking can guarantee that the design is always correct in all possible application scenarios.
However, due to the state explosion problem, the checking cost is exponential to the scale of the model.
We have to limit the scale to achieve practical checking.

We argue that the small scale model checking can still find most deep bugs.
Our argument is backed by the observation known as the \textit{small scope hypothesis} \cite{Jackson12, Yuan14, Maric17, Hance21}.
The hypothesis states that analyzing small system instances suffices in practice to ensure sufficient correctness of large scale systems. Empirical studies support this hypothesis in different settings \cite{Andoni03, Oetsch12, Yuan14}.
For example in the setting of distributed systems, this hypothesis is proved for a family of consensus algorithms targeting the benign asynchronous setting \cite{Maric17}.
An empirical study \cite{Yuan14} of 198 bug reports for several popular distributed systems found that 98\% of those bugs could be triggered by three or fewer processes.

We leverage our understanding of the CRDT protocol to tune the key parameters of the system configuration in order to control the scale of the model.
Let $n$ denote the number of server replicas, $q$ the number of client requests and $Q$ the number of all possible client requests, i.e. the size of the space of all possible requests.

As for the server number $n$, we set $n$ to a small number. In our application of the \met framework (Section \ref{Sec: Exp}), $n$ is up to 3. 
%In many cases 3 is an appropriate number for the number of replicas.
In most cases, 3 replicas are sufficient to manifest the subtle interleavings of concurrent data updates which result in the deep bugs \cite{Sun14, Xu14}.
As $n$ goes beyond 3, the checking cost quickly becomes prohibitive.

As for $q$, we set the $q\geq n$. This is to make sure that every replica is covered. 
Besides, the client requests also need to cover all APIs the CRDT provides.
Given the limit on the testing budget, we can have bigger $q$ when $n$ is small, but when $n$ is relatively large, we can only have $q$ close to or equal to $n$.

As for $Q$, we let it be a little larger than $q$. This enables each client request to possibly have different parameters while limiting the state space.
The value of $Q$ is affected by multiple factors.
The \met framework is mainly targeted at data container types.
We first need to consider whether the data elements in the container are independent. If so, we can consider each element separately, significantly reducing the size of the state space. 
But we may also face the data types where the data elements cannot be handled separately.
For example elements in the \rwff-RPQ can be viewed as independent\footnote{When considering the conflict resolution, the elements in the \rwff-RPQ can be considered independent. When the server replica locally maintains the priority queue data structure, the data elements are dependent. But this dependence is not relevant to the conflict resolution we focus on in this work.}, while elements in the \rwff-List are not independent. % (See Section \ref{SubSubSec: Threats-Pruning}).
When the model checker needs to execute a client request, we set the parameters of the request by choosing a uniformly random request from all possible requests.
When the operations have multiple equally important parameters, we choose one to represent all the parameters. 

%We tune all the factors above to let $Q$ be a little larger than $q$. %, to let each one of the $q$ requests can choose a different
%
Detailed configurations of the model and the results of model checking are described in Section \ref{SubSubSec: Threats-Pruning}.

%--
%--
\section{Explorative Testing of CRDT Implementations} \label{Sec: Code-Level}

The \met framework provides an explorative testing service to guarantee sufficient correctness of CRDT implementations.
The key of the explorative testing service is to utilize the model checking results at the model level, and guide the explorative and exhaustive (constrained by the testing budget) testing at the code level.
The two key components of the testing service are: automatic \textit{test case generation} from the model checking trace and \textit{testability enhancement} of the system under test, as detailed below.

%--------------------------------------------------------------------------------
\subsection{Test Case Generation} \label{SubSec: Test-Case-Gen}

\begin{figure}[htbp]
    \centering
    \includegraphics[width=\linewidth]{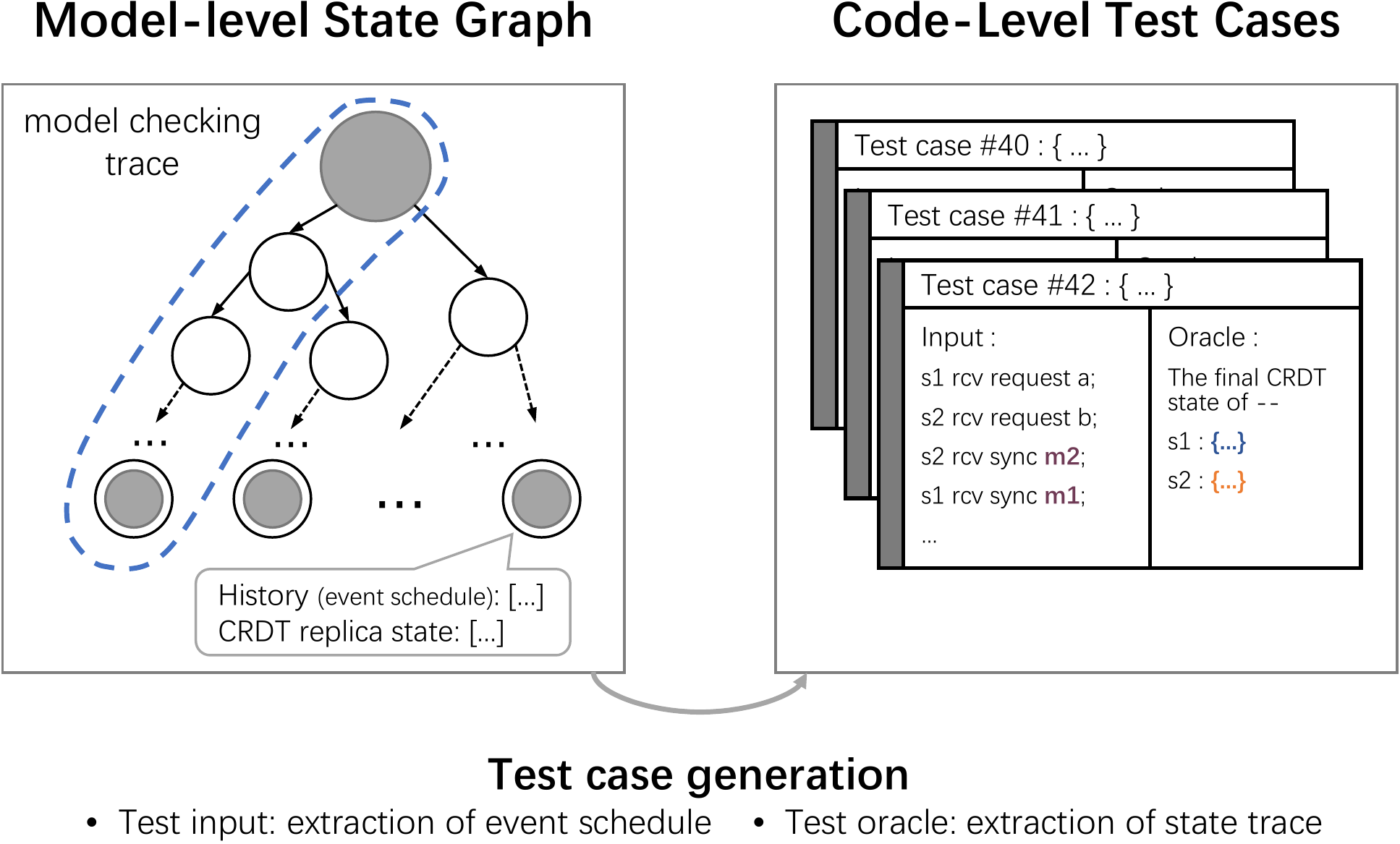}
    \caption{From model-level traces to code-level test cases.}
    \label{F: Test-Case}
\end{figure}

The salient feature of explorative testing is that it systematically controls and traverses the non-deterministic choices of message reorderings for CRDT implementations.
Given that the exhaustive checking of message reorderings has been conducted in the model layer, we automatically generate test cases from the model checking traces.
The test case generation consists of two parts: the generation of test inputs and that of the test oracle, as shown in Figure \ref{F: Test-Case}.

%--

%----------------------------------------
\subsubsection{Generation of Test Inputs} \label{SubSubSec: Gen-Input}

The test input is merely the \textit{event schedule}, i.e., a sequence of events that take place in the system.
The test input consists of the \textit{system events} and the \textit{environment events}.
According to our modeling in Section \ref{SubSec: TLA-Spec}, the system events mainly include data update requests from the clients and the data synchronization requests among replicas.
The environment events are the delivery of data updates from the network channel to the replicas.
%According to our modeling in Section \ref{SubSec: TLA-Spec}, the \textsf{do} and \textsf{send} events are system events, while the \textsf{receive} events are environment environments.

To enable automatic extraction of test inputs %, i.e. all the system and environment events discussed above,
from the model checking trace, we add the history variable \cite{Lamport2017Auxiliary} in the TLA+ specification of the CRDT protocol. 
The history variable is an auxiliary variable which passively records the state transitions in system execution without interfering with the execution.
%to the model to ``piggyback" the event sequence in the state. 
The test inputs can thus be extracted from the history variable.
%Thus, we can get all possible event sequences from the history variable of each state.

%The testing controller will read the system events and conduct the tests according to the events.

%\hy{代码层，几个术语要敲定。光是testability不够用。是不是叫test manager？预先明确说明清楚即可。}

%In design of a CRDT protocol, we do not consider the crash of server replicas or the loss of messages, as discussed in Section \ref{SubSec: TLA-Spec}.
%Thus we do not need to generate environment events like node crash or message loss.
%The main uncertainty we consider is the reordering of messages. The permutation of message orders has been considered in the model checking of CRDT design.

%----------------------------------------
\subsubsection{Generation of the Test Oracle}

In \mett, the executable code and the TLA+ specification of the CRDT protocol are viewed as just two steps on a ladder of abstraction. 
Given that the protocol design has been verified by model checking, our objective is to verify that 
the code-level execution is correct with respect to the model-level execution.
%the code is correct with respect to the specification.
%
Thus, the sequence of system states in every model checking trace is extracted to serve as the test oracle.

In explorative testing, we pick one test case and control the system execution to follow the event schedule in the test input.
%The execution requires enhancement of the testability of the system under test, as detailed in Section \ref{SubSec: Testability}.
%We log the system execution trace, i.e., the sequence of system states.
We check whether the code-level sequence of system states is equivalent to the model-level one.
Here, the criteria of being equivalent is that, the meta-data for conflict resolution and the concrete data of the data type are the same, while certain details in the code-level states can be ignored.
%

%The test oracle consists of a sequence of (model-level) system states, starting from the initial state to the ending state. 
%The system execution trace is required to be equivalent to the states of the model-level trace. Here ``being equivalent" is defined as ``having the same key CRDT data".
%Theoretically if we can check the states in the code-level trace to be equivalent with its corresponding model-level state, we can prove the implementation is correct w.r.t. this trace. If it holds for all the traces, we can conclude that our CRDT implementation is correct w.r.t. the model checking restrictions we set.

%\vspace{6pt}
%In summary, we obtain test cases from the final states of model-level execution traces.
%The test inputs are the sequences of system events obtained by the history variable of the final states.
%And the test oracles are the key CRDT data of the final states.

%--------------------------------------------------------------------------------
\subsection{Testability Enhancement} \label{SubSec: Testability}

\begin{figure}[htbp]
    \centering
    \includegraphics[width=\linewidth]{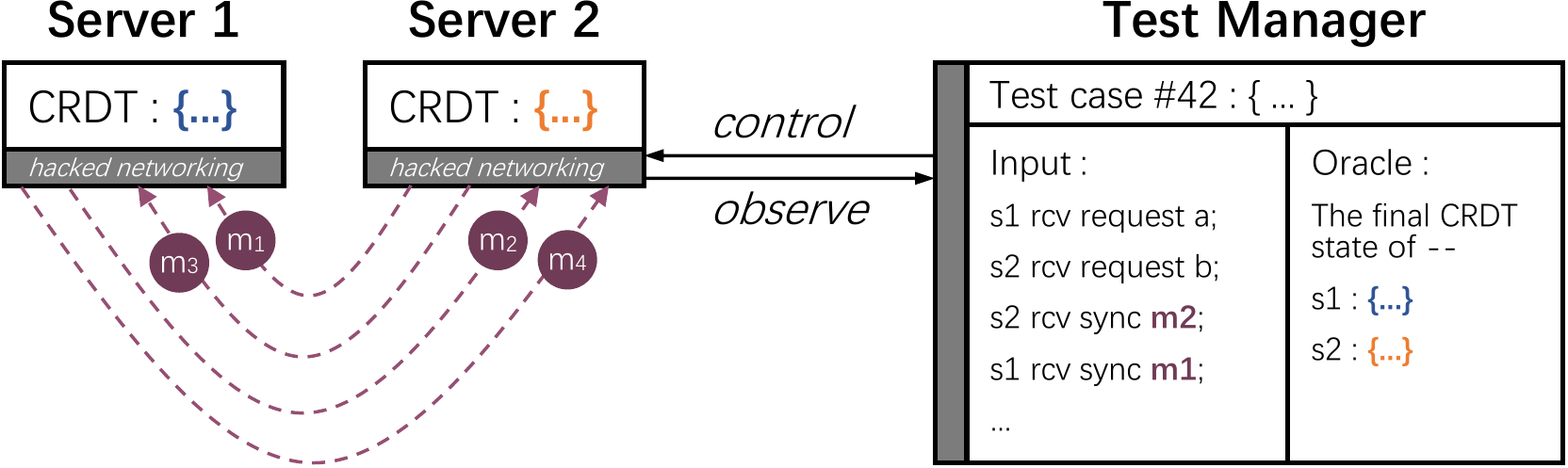}
    \caption{System controllability and observability.}
    \label{F: Testability}
\end{figure}

The critical challenge in explorative testing is to control the system execution to deterministically follow the event schedule in the test input.
Moreover, the test manager also needs to inspect the internal state of the server replica, in order to tell whether the system passes certain test cases.
The rationale of testability enhancement in \met is shown in Figure \ref{F: Testability}.
We discuss the two primary issues in testability enhancement in detail below.

%--

%----------------------------------------
\subsubsection{Controllability: Manipulating Code-Level Execution}

In TLA+ we model a distributed system in terms of one single global state. 
Thus the event schedule in the model checking trace is a total order of events.
%The event schedule is extracted as the test input.
%The state transition events are extracted from the model checking trace and serve as the test input.
%
The test manager needs to deterministically control the system execution to follow this total order of events. %the event schedule in the test input.
%Now working with a running system, the test manager needs to control the execution of the system to follow the state transition events in the test input.
%To achieve this goal, we need to enhance the controllability of the system under test.

The key to achieving system controllability is to control the timing of each event.
In the \met framework, we hack the networking component of each server replica.
The hacking is transparent to the upper layer protocol. 
The replicas are provided with the illusion that they communicate with each other via sending and receiving messages.
In fact, all the sending and receiving of messages are redirected to the central test manager.
The test manager is in full charge of deciding the timing of every event. %, and controlling the system execution.
The system execution makes one step forward when the test manager "fires" one event.

Given the system architecture explained above, we directly integrate the client request generation module into the test manager, and do not explicitly model the clients.
The test manager will dispatch client request events to the replicas as specified in the test input.

%The principle provided in \met is applicable to different systems. 
In our application of \mett, we instantiate the general approach above over the CRDT-Redis data store \cite{CRDT-Redis}.
%experimental evaluation of \mett, we apply this principle to enhance the controllability of the CRDT-Redis data type store. 
Details of our implementation can be found in Section \ref{SubSec: MET-Impl}.

\subsubsection{Observability: Inspecting Internal State of the Replica}

%When model checking, the detailed state of the model is visible to the model checker. It checks the specification with such detailed information of the state. And we further use the model state as the test oracle. 

Though we can obtain part of the replica state via its public API, the \met framework requires much more than that.
As indicated by the test oracle, the test manager requires detailed metadata for conflict resolution. 
It also requires the internal representation of the data structure.
%Given the test oracle, all info related to the replica state as specified in the test oracle needs to be observed.

The commonly used approach is to obtain the required data through extensive logging. However, this logging will require a detailed description of the timing information. It then requires a non-trivial offline analysis of the log. %of system execution.
%
%Given that the test manager fully controls the execution of the system, 
The \met framework requires the data type store to be enhanced with replica state inspection APIs. 
Such APIs are dedicated to the test manager whenever it needs to inspect and record the internal state of the server replica.

Given that any progress of the system execution is under the control of the test manager, the test manager can use the APIs for replica state inspection and obtain the system state after every event is dispatched to the replicas.
Thus the state sequence of system execution directly corresponds to the test oracle.
%
%According to the discussion above, the approach \met adopts to enhance system observability 
This approach greatly eases the conformance checking against the test oracle.

%It is needed to further increase the observability of the system. This will enable us to check whether the code-level execution satisfies the test oracle.
%Therefore we developed additional read operations in the CRDT-Redis to get the raw state of the CRDTs in the server. With these raw read operations we can compare the final states between the test cases and the system.

%However, usually we can not get all the detailed data and metadata of the server in the real world.
%In our case of CRDT-Redis, we can only read the interpreted data of the CRDT, such as the max element in the RPQ, and the order of the existing elements in the list. We can not get the raw data and metadata of the CRDT. For example, the vector clock, the pid and the lamport clock for concurrent conflict resolution, and the innate / acquired value\cite{RWF-CRDT} to compute the true value of the elements.

%--
%--
\section{Application of \met in Practical CRDT Development} \label{Sec: Exp}

In this section, we present how we apply the \met framework to cope with deep bugs in our CRDT design and implementation. %(as discussed in Section \ref{SubSec: Expe-Moti}).
We first present the implementation of \mett. 
Then we discuss how the deep bugs are found and fixed in both the model layer and the code layer.
In the end, we discuss how \met improves our confidence in the correctness of our CRDT design and implementation.

%--------------------------------------------------------------------------------
\subsection{Implementation of the \met Framework} \label{SubSec: MET-Impl}

We specify our CRDT protocols in PlusCal \cite{Lamport09}, which is automatically translated to TLA+ specifications.
The \rwff-RPQ protocol has about 200 lines of PlusCal code, and \rwff-List about 300.
%We check whether the protocols guarantee eventual consistency. We also checker constraints on the position identifier for the RList. 
%
%the strong eventual consistency of the models, which means that each two replicas should have the same state if they have executed / synchronized the same set of operations.
%Here having the same state means that two replicas has the same data type data and the same metadata for conflict resolution.
%
We model check and revise the design until no violations can be detected.
Every model checking trace of the final design is parsed to generate the test case.
The test case generation tool has about 100 lines of Python code. 
%for parsing the set of final states into the test cases. Each test case contains the event sequence as test input, and the expected final state data and metadata as test oracle.

%extracted as the test oracle. 
%In practice, we just compare the final state.
%The TLA+ model checker will output a set of final states of each round of model checking. The final state means that all the client requests are handled and all the synchronizations are finished. 

%The event sequences that generate each state are in the history variables. The event sequences are extracted as the test input.

%\mybreak
    
% \zyq{observability、log部分要写两句，保持形式完整。} 

%\hy{找到首次RWF的地方，要对RWF作解释，不能简单指向ref。}
    
Our implementation of the \rwff-RPQ and \rwff-List are over the CRDT-Redis data type store we developed \cite{CRDT-Redis}.
To enhance the testability of CRDT-Redis, we first hack the networking module of the server replica.
%All communications between replicas are redirected to the central test manager.
%Note that the hacking of the network module is transparent to the upper-layer CRDT protocol.
Then we augment each data type with APIs dedicated to the test manager for replica state inspection.
The testability enhancement is implemented in about 200 lines of C code.

%we hacked it with less than 200 lines of C code. Most of them are focused on the P2P replication part (controllability) and the CRDT read operation parts (observability). 

%We added a test mode to the P2P replication component. It will make the server believe that it is in a replication cluster with many other server nodes, while it is only connected to the controller node. 

%And we also add read operations for RWF-RPQ and RWF-List implementations to read their metadata.

% We made a small change to the RWF-List implementation. Originally the RWF-List implementation introduces a random number when inserting a new element. This is an optimization that may reduce the memory overhead of the metadata of the list. We removed this random number for the deterministic execution that is instructed by the test cases. We are confident that this will not hide any potential problems.

The test manager is implemented in about 1000 lines of C++ code.
The test manager is in charge of launching the server replica groups.
%It will start several Redis servers in the P2P test mode. 
It will intercept all communications among the servers and strictly control the system execution according to the test input.
%the server synchronization messages. Furthermore, it will read the test cases and then send client requests / forward server synchronization messages accordingly. 
During the (controlled) system execution, the test manager inspects the internal states of the replicas at its will.
In our practice, we just inspect the final states of the replicas when all messages are delivered to reduce the testing cost.
%Finally it will read the CRDT states from servers and check if they match the final states in the test cases.
%
%In practice, we do some optimization. We only check the final state of the traces. 
%This greatly reduces our workload to test the system. 
This simplification is reasonable because in the system state, we record all the metadata for conflict resolution and all data of the data type itself.
Our experiences in CRDT development convince us that, any divergence in the previous states will be reflected in the final state, and the divergence will be amplified by more data updates.
The probability that the divergence is counteracted by following updates is negligible.

%\sucai{Divergence in the complex metadata will be not be counteracted by consequent updates. On the contrary, they will diverge even more significantly. 
%The probability for the code-level trace being equivalent with the model-level trace is very high if the final states are equivalent.
%}

%--------------------------------------------------------------------------------
%\subsection{Execution Environment for \mett}

%实验的各类配置。整理、阐述。

The experiment is conducted on a PC with an Intel I7-6700 quad-core CPU (3.40GHz) and 16GB RAM. %, running Windows 10 Enterprise v2004.
We use VMware Workstation 16 Pro to run the virtual machines. Each VM is set to have a quad-core CPU with 8GB RAM, running Manjaro with Linux kernel 5.10.
The model checking is conducted with TLA+ Toolbox v1.7.1 \cite{TLA-Toolbox}. 
%The testing is conducted in the virtual machine.
%
All implementations discussed above are available in our GitHub repository online \cite{CRDT-Redis}.

With the help from \mett, we find and fix seven bugs in our CRDT design and implementation. Note that we mainly focus on relatively deep bugs here. Bugs which can be detected by standard testing techniques are omitted.
The bugs found in our application of \met are listed as Issue \ref{I: Iss1} \textasciitilde \ref{I: Iss7} in the issue page of the online repository \cite{CRDT-Redis}.
As discussed in Section \ref{SubSubSec: Why-MCK}, Issue \ref{I: Iss1} can be found by stress testing, but with \mett, we find and fix it at a much lower cost.
When we use \met to fix Issue \ref{I: Iss1}, we further raise Issue \ref{I: Iss3}, which requires the revision of our CRDT design.
Issue \ref{I: Iss2} can be detected by manual explorative testing. 
\met principally automate this process, at a much lower cost.
For Issue \ref{I: Iss4} \textasciitilde \ \ref{I: Iss7}, they can only be detected by \mett.
We discuss these bugs in detail below.

%--------------------------------------------------------------------------------
\subsection{Model Checking of Our CRDT Design}

We detect Issue \ref{I: Iss1} by hours of stress testing.
We then take days to fix the issue.
%With \met we first specify the protocol in TLA+.
%Then we model check the design (see detailed model checking statistics in Section \ref{SubSec: Met-Confidence} below).
%
Using the \met framework, we first precisely specify the CRDT protocol in TLA+. 
%Since the conflict resolution in the CRDT protocol often involves tedious enumeration of quite a number of conflict patterns, 
The formal specification and model checking help us find bugs deep in subtle corner cases in our design.
The fixing of Issue \ref{I: Iss1} using \met is much more efficient. 
The model checking provides the trace which leads to the violation of the correctness condition.
This is the shortest trace since the TLC model checker traverses the state space in the BFS manner. 
We use this trace to help revise the design.
We keep revising the design until the design passes the model checking.

The bugs found by the model checking have the following characteristics.
Such bugs can only be triggered by specific interleavings of client requests and replica synchronization operations. 
The bug-triggering event pattern may not be quite long, but it hides in an exponential number of all possible such patterns.
More importantly, the bugs are often latent, in the sense that they may not cause externally observable symptoms for quite a long period of time even after the divergence has appeared.
%
%Thus, such bugs are often out of reach of human reasoning. 
Thus finding such bugs often requires long time random stress testing. 
All these characteristics of the bugs necessitate the formal specification and verification of our CRDT design. %an effective and efficient approach.

The formal specification process in the fixing of Issue \ref{I: Iss1} drives us to further improve our design, as shown in Issue \ref{I: Iss3}.
In our original design, we treat the \readd operation as a special case in the $add$ operation.
The semantics of the \textit{add} operation becomes quite complex.
Issue \ref{I: Iss3} basically says that the design should be revised.
The \readd logic should be treated as a separate operation.
The conflict handling logic involves the new operation \readd should be supplemented.
Though we put more efforts in designing the new operation \readd as well as the conflict resolution logic involving \readd, the overall complexity of our CRDT design is reduced and the resulting specification is less complex. 
Separating the \readd operation out also helps reduce the model checking cost.

%--
\begin{issue}{3} \label{I: Iss3}
    The \readd operation should be separated from the \textit{add} operation and treated as an independent operation \rm{\cite{Issue3}}.
\end{issue}

Summarizing the process of fixing Issue \ref{I: Iss1} and \ref{I: Iss3}, the formal specification forced us to eliminate the ambiguity in semantics of element existence corresponding to the \readd operation.
%model the existence of elements corresponding to the \readd operation with finer granularity.
We explicitly model the existence of one element into three cases: \textit{existent}, \textit{non-existent} and \textit{once-existent}.
The meanings of the "existent" and "non-existent" states are as usual.
The state "once-existent" pertains to the \readd operation.
When we \readd an element "a", this element should not be treated as a new element.
The \readd operation adds an element which once existed in the data container but is now not in the container.
We model this status as "once-existent" to correctly design the semantics of the \readd operation.

%--------------------------------------------------------------------------------
\subsection{Model Checking-driven Explorative Testing of Our CRDT Implementation}

Having verified the CRDT protocol in the model layer, we further conduct model checking-driven explorative testing in the code layer.
We aim at finding bugs due to transcription errors, named \textit{transcription bugs}, and bugs due to assumptions made in the protocol, named \textit{assumption bugs}.

%----------------------------------------
\subsubsection{Transcription Bugs} \label{SubSubSec: Trans-Bug}

We use Issue \ref{I: Iss4} as an example of transcription bugs in Section \ref{Sec: Moti-Ovw}.
This bug is beyond the coverage of stress testing, and can only be found by \mett.
When transforming the design, both formal and informal, into codes, the developers inevitably needs to handle details which are not covered in the design.
Issue \ref{I: Iss4} is a bug caused by the deviation between our design and implementation, corresponding to the "dummy-existence" semantics of data elements.
%
%The \rwf CRDTs are designed using the RWF framework.
%Here we have \rwf-RPQ and \rwf-List. 
%%
%When implementing them, we implement the RWF framework first. 
%Then all the RWF CRDTs can be implemented with it. 
%With such intention, we extracted the commonality of the RWF RPQ and RWF List designs.
%The commonality we extracted is the "existence" of elements. It indicates whether an element should be considered as exist in the CRDT.  
%

Specifically, at the design level, the element is either exiting or not-existing. The handling is protected by the precondition that ``the element must exist".
This implicitly states that, if the element does not exist, nothing should be done.
%
%In the code level, the existence must have three cases. The element may be not existing in the container.
%The element may normally exist in the container.
However, at the code level, we must model the intermediate state that the element exists as a dummy.
This intermediate state is mainly caused by the late arrival of an operation adding some element into the data container.
%
%For example in a RPQ, the element is first added with an initial priority and then its priority is increased by 10.
%However, a replica can possibly receive the update operation first and later the insert operation.
%Upon receiving the update operation, the element can not be correctly handled. It is a dummy.
%The update operation can either be rejected or be buffered to be handled later when the insert arrives. 
%This depends on the user requirement.
%
In our buggy implementation, the dummy state is not explicitly modeled and illegal positions are assigned to the dummy elements (see details of the example in Section \ref{SubSubSec: Why-Testing}).

Issue \#5 \cite{Issue5} and Issue \#6 \cite{Issue6} are similar to Issue \ref{I: Iss4} in the sense that they are also due to the misinterpretation of the design, concerning the existence of elements.
We omit the detailed discussions on these two issues due to the limit of space.

%----------------------------------------
\subsubsection{Assumption Bugs} \label{SubSubSec: Assum-Bugs}

The assumption bug Issue \ref{I: Iss2} is introduced when we integrate different types of CRDT protocols in the CRDT-Redis platform.
Different protocols have different assumptions about the underlying network.
There is often a gap between what the protocol assumes and what the platform provides.
This significantly increases the odds of assumption bugs in our CRDT implementations.
Moreover, implementing a patch fixing such bugs is also error-prone and needs extensive testing (see details of this issue in Section \ref{SubSubSec: Why-Testing}). 

%In the \rwf framework, we do not need causal delivery.
%The CRDTs using the RemoveWin strategy, RemoveWin-List and Re-moveWin-RPQ in our case, need causal delivery.
%We overlook this difference in assumption.
%We do not implement causal delivery of messages for RemoveWin CRDTs.
%
%We implement a patch to fix this bug. 
%This patch basically buffer messages. 
%One message is delivered from the buffer to the upper-layer protocol when all its causal predecessors have already been delivered.
%This patch is moderately complex.
%%In our first implementation, server$_2$ synchronizes $b$ for $a$, but misses $c$.
%%Then when server$_2$ receives $d$, it keeps waiting for $c$ and then get stuck.
%We introduce bugs when fixing the bug in Issue \ref{I: Iss4}.
%The implementation of this patch also needs extensive testing.

Issue \ref{I: Iss7} is another example of an assumption bug. The \rwff-List and the RemoveWin-List protocols assume that there is a position ID generator. It generates list element IDs which are totally ordered, dense, and consistent with the element order resulting from the $add$ operations. 
Our design of list element organization (borrowed from the existing design in \cite{Weiss10}) does not specify how such an ID generator can be implemented. 
The implementation of such an ID generator is non-trivial, and the bug in our implementation of the ID generator is found by \mett.
%The requirement on the ID generator is reflected in our TLA+ specification of the correctness condition for the replicated list data type.
%Given the error trace provided by \mett, we find the root cause and fix the bug.
%
Given the test case execution trace, as well as how the trace violates the test oracle, we quickly find the root cause and revise our implementation of the ID generator.

%--
\begin{issue}{7}\label{I: Iss7}
    The order of elements in the RemoveWin List and the \rwff-List is sometimes not consistent with the order induced by $add$ operations \rm{\cite{Issue7}}.
\end{issue}

%--------------------------------------------------------------------------------
\subsection{How \met Improves Our Confidence in CRDT Implementations} \label{SubSec: Met-Confi}

Our primary goal of introducing the \met framework is to obtain correct CRDT designs and implementations.
In the ideal case, the design has been verified to be correct by model checking.
As for the implementation, the testing is both explorative and exhaustive, driven by model checking.
Thus the implementation is also correct.

However, in practice, the design and implementation obtained using \met still potentially have bugs.
There are mainly two sources of threats to the correctness of CRDT designs and implementations ensured by \mett.
One is that any details in the implementation which are not modeled in the formal design can still cause bugs.
The other is that the state space is not really traversed. 
A significant portion of the state space is pruned due to the state space explosion problem.
This pruning can potentially miss bugs in the design and implementation.

We discuss in detail below how \met provides us with sufficient correctness of CRDT designs and implementations, in spite of the threats discussed above.

%----------------------------------------
\subsubsection{Threats from System Modeling}

There are always certain aspects of a system operating in realistic scenarios which are not modeled in (both informal and formal) system design.
Thus bugs corresponding to the unmodeled part of the system cannot be detected by \mett.
The unmodeled part of the system is principally the "unknown unknowns" \cite{Garlan10, Esfahani11, Elbaum14}.
We argue that such bugs will not hamper the usefulness of \mett.
\met is mainly targeted at the subtle deep bugs in CRDT design and implementations.
The deep bugs we focus on are primarily "known unknowns" \cite{Garlan10, Esfahani11, Elbaum14} as detailed below.

%
%The modeling process inevitably omit certain details which are present in the implementation.
%The details omitted in the modeling will not be covered by the formal specification and verification.
%Nor will they be covered by the model checking-driven exploartive testing.
%This means that such details, as well as the bugs pertaining to them, cannot be detected by \mett.
%

According to our experiences in CRDT design and implementation, deep bugs are mainly known unknowns in the sense that we know or highly suspect their existence. 
We also know the high-level pattern and mechanism of how the deep bugs may manifest themselves.
In theory, the developer can find and fix all the deep bugs through manual reasoning and testing.
This approach just requires too much human efforts and time.
We lack the tool support to automate and accelerate this manual process, effectively reducing the testing cost.

Our \met framework fills this gap.
The gist of proposing the \met framework is thus to automate the most tedious and repetitive part of human reasoning and manual explorative testing.
Given the automation provided by \mett, human knowledge of the deep bugs can be efficiently leveraged, to dig out the deep bugs at a reasonable cost.

In summary, though \met is no better than the human knowledge of deep bugs, it efficiently leverages knowledge of the known unknowns and pragmatically finds and fixes deep bugs.
In this regard, \met can provide us with sufficient confidence in the correctness of CRDT designs and implementations. % in the sense that, the number of deep bugs are significantly reduced.

%Such deep bugs are out of the reach of standard testing techniques.

%For example, the coarsening of the specification intentionally omit certain details (see Section \ref{SubSubSec: Coar-Spec}).
%We leverage the designer's understanding of the CRDT protocol to ensure that the coarsening will not miss deep bugs.

%When we have more insights into the design and implementation of CRDTs, the \met framework can still leverage these insights to achieve better modeling and testing.

%----------------------------------------
\subsubsection{Threats from State Space Pruning} \label{SubSubSec: Threats-Pruning}

The total cost for model checking-driven explorative testing is mainly decided by the cost for the model checking.
It is because the cost for the explorative testing is proportional to the number of test cases, which is the same as the number of model checking traces.
We have to limit the scale of the model to finish the testing within the testing budget.
The limit on the model scale is based on the small scope hypothesis, as discussed in Section \ref{SubSec: MCK}.
%After limiting the model, we can complete the model checking and the explorative testing. 
Table \ref{T: MET} shows the different parts of the cost for testing via the \met framework.
The configurations shown in the table are the largest affordable scale of models.
The \met configuration tuple stands for "number of servers -- number of client requests".
In our experiment, we limit the testing time by 12 hours and limit the disk space used for storing the model checking traces by 100GB.

%--
\begin{table}[htbp]
    \centering
    \caption{The cost for explorative testing using \mett.}
    \renewcommand{\tabularxcolumn}[1]{m{#1}}
    \newcolumntype{H}[1]{>{\hsize=#1\hsize\raggedright\arraybackslash}X}
    
    \begin{tabularx} {\linewidth} {H{0.52}H{0.94}H{1.26}*{4}{H{1.07}}}
        \toprule
        &&&\multicolumn{2}{l}{Space cost}&\multicolumn{2}{l}{Time cost}\\
        \cmidrule(r){4-5}\cmidrule{6-7}
        &\met config&Num of traces&TLC output&Test case&\small{Checking}&\small{Testing}\\
        \midrule
        \multirow{3}{=}{\rwff-RPQ}
        % &2-4&\num{150164}&32MB & 15MB &15s&2:48s\\
        &1-10&\num{9765625} &2.87GB&1.22GB&6:30s&3:10:46s\\
        &2-5&\num{7202312} &1.80GB & 884MB & 9:26s& 2:29:03s\\
        &3-3&\num{883413}&209MB & 105MB &1:53s&20:27s\\
        \cmidrule{1-7}
        \multirow{3}{=}{\rwff-List}
        &1-5&\num{69343957} & 9.64GB & 5.47GB & 2:43:22s & 11:50:55s\\
        &2-4&\num{23639180}&4.09GB & 2.1GB &2:12:38s&5:09:29s\\
        &3-3&\num{5085147}&1.43GB & 731MB &51:05s&1:46:18s\\
        \bottomrule
    \end{tabularx}
    
    \label{T: MET}
\end{table}

The model checking will not complete if we increase any dimension of the configuration. For example, we tried to increase one client request to the RWF-RPQ model configuration 3-3, which is 3 servers + 4 client requests. 
The model checking will not finish after 24 hours. Looking at the number of pending states in the queue, we estimate that it needed at least another 24 hours. It has used about 150GB of disk space when we manually stopped the checking.

In the model checking, we have at most three server replicas.
The case of one server is used to exercise the basic logic of the protocol without any concurrent updates.
The case of two servers is used to test the commutative properties of possible concurrent operations.
The case of three servers is used to test the subtle corner cases in conflict resolution. 
As proved in conflict resolution with the operational transformation technique in collaborative editing, three operations are enough to express all the properties required for eventual convergence \cite{Xu14, Sun14}.
Thus we think that the model checking with three replicas %Although we can only exhaustively do our MBT under such limits, we think it 
can find most deep bugs pertaining to the conflict resolution.
Also note that we have three types of operations for each CRDT. 
%The CRDTs under testing have no more than three data update APIs.
Thus we have at least three client requests for each test case, where all types of CRDT operations can be tested.

As shown in Table \ref{T: MET}, the cost for \rwff-RPQ and that for \rwff-List differ quite a lot. 
This is mainly due to the fact that the \rwff-RPQ is data-independent, while the \rwff-List is not.
Our checking for the \rwff-RPQ focuses on one data element. 
We choose \{10, 20\} to be the set of possible initial values of $add$ operations, and \{-3, 4\} the set of possible value increases for $increase$ operations. 
All values are picked uniformly at random.
%This value spaces makes the value of element distinct within about 10 operations.
The \rwff-List does not have the property of data independence.
The order of the elements is important for the list. 
Since each client request may add a new element to the list, we set the size of the element space to the total number of client requests. 
As we target at collaborative editing scenarios, each data element can have up to 6 attributes (e.g., the font, being italic and being bold face).
In our testing via \mett, we consider only one attribute since we mainly focus on the conflict resolution logic.
We choose \{10, 20\} to be the set of possible initial values and update values for the \rwff-List.

%【空间】

We also propose an optimization concerning the model checking trace.
We did not let the TLC model checker output the whole state transition graph like the work in \cite{Davis20vldb}.
Instead we let the CRDT models output only the final states. 
This makes the output file one or two orders of magnitude smaller.
%This makes the output file dozens of times smaller.
The extracted test case files are about half the size of the TLC output files. %, which is smaller.
The optimization prevents the space usage from being the bottleneck.
%Now we find that we do not need to worry about the space cost of our MBT. 
%The state explosion problem manifests itself in terms of the time cost. 
%
%For example the time cost for testing. 
To get a rough idea, it takes about one hour to test 1GB of test case data on average.
%Thus our disk space can support model checking of hundreds of hours of time.

%【时间上】

Comparing the time cost for the model checking and that for the explorative testing, we find that the model checking is faster than the explorative testing. 
This is mainly because the model checker can prune the redundant checking of system states, while in \met execution of each test case is stateless.

According to the discussions above, the cost for testing via \met is reasonable. 
Now the critical issue is to discuss whether \met can help us effectively find deep bugs in our CRDT design and implementation.
The effectiveness of the \met framework depends on the small scope hypothesis.
We argue that the hypothesis works well in the testing of CRDT implementations.
In collaborative editing scenarios, convergence properties involving at most three concurrent operations are sufficient to ensure eventual convergence.
Since operational transformation techniques also focus on ensuring eventual convergence among replicas, this result indicates that with three replicas we can construct the triggering pattern of most deep bugs in CRDT development.
This result is in accordance with our experiences in CRDT testing.
As discussed in our design of \mett, we focus on the conflict resolution logic of CRDTs.
We leverage our understanding of the CRDT design to carefully prune parts of the state space which are not relevant or less relevant to the conflict resolution logic.

%Given that the system can be limited to a small scale, the \met framework completes model checking and explorative testing in reasonable time.

In summary, most deep bugs concerning the conflict resolution logic can be reproduced in small scale systems.
The pruning in our model checking in \met carefully avoids erroneously pruning traces which can trigger deep bugs.
In this regard, we believe that \met ensures sufficient correctness of CRDT designs and implementations.

\section{Related Work} \label{Sec: RW}

Distributed systems are notoriously difficult to implement correctly, mainly due to the intrinsic uncertainty in their executions \cite{Fonseca17}.
This uncertainty is created by asynchrony, the absence of global time, various types of failures, etc.
From the perspective of adversary argument \cite{Jeff-Erickson-Notes}, the developer needs to enumerate all possible combinations of uncertain events and reason out the invariable correctness of system behavior.
This type of exhaustive reasoning is often impractical for the developer.
This explains why model checking techniques are often employed to verify the correctness of distributed system designs \cite{Newcombe11, Newcombe15}.
For example, the Chord protocol is formally specified in Alloy and is model-checked using the Alloy Analyzer \cite{Zave15, Zave17}.
Distributed consensus protocols, such as Paxos \cite{Paxos-TLA} and Raft \cite{Raft-TLA} are specified in TLA+ and model-checked using the TLC model checker.
TLA+ is also widely used to verify the design of distributed systems in industrial scenarios, e.g. in Amazon S3 and DynamoDB \cite{Newcombe14, Newcombe15}, CosmosDB \cite{CosmosDB-TLA} and TaurusDB \cite{Gao21}.
The techniques discussed above mainly focus on design at the model level. From a pragmatic perspective, we need to leverage model checking to further improve the reliability of code-level implementations.

Distributed System Model Checking (DMCK) treats the code-level implementation as the model, and directly conducts model checking on the code \cite{Musuvathi02, Yang09, Guo11, Lee14}.
DMCK precisely controls the execution of the distributed system and exhaustively explores all possible execution paths.
DMCK has shown the potential for detecting deep bugs which are extremely difficult to detect through traditional testing techniques. 
However, DMCK usually imposes a prohibitive cost, and state reduction techniques are essential to the practical application of DMCK.
Partial order reduction exploits the independence among events, and establishes the equivalence among interleavings of independent events \cite{Yang09}.
Thus the state space is reduced by checking only one interleaving on behalf of all equivalent ones.
Dynamic interface reduction \cite{Guo11} is based on the observation that different interleavings of local events (within one system component) often result in the same global interaction (among multiple components).
Thus the redundant enumeration of global events can be saved.
To better reduce the state space, semantic knowledge of the target system can further be utilized.
The Semantic-Aware Model Checking (SAMC) \cite{Lee14} presents four semantic-aware reduction policies that enable the model checker to define fine-grained event dependency/independency and symmetry beyond what black-box approaches can do.
The Failure Testing Service (\textsc{Fate}) focuses on recovery testing \cite{Gunawi11}. 
\textsc{Fate}  exhaustively exercises as many combinations of failures as possible.
Semantic knowledge of the failure patterns is utilized to prioritize the failure patterns. 
The limited testing budget is then directed to the testing of the high-priority patterns.

Our \met framework shares the same goal of DMCK techniques. 
We all aim to inherit the feature of exhaustive exploration of model checking to improve code-level testing.
%Though fully exhaustive testing in the code level is usually not practical, we aim at obtaining sufficient correctness at reasonable cost. 
However, the \met framework takes a methodologically different approach.
\met does not aim at fully automatic checking of the code-level implementation as the model.
\met is designed to be semi-automatic. 
Human efforts are required to obtain the formal specification of the system design.
The developer is also required to enhance the testability of the system under test.
After the manual preparations above, the model checking, the test case generation and the explorative testing are automatic.

We opt for this semi-automatic approach mainly based on the observation that, fully automatic DMCK approaches still imposes a prohibitive cost on distributed systems in the field. 
We leverage moderate human efforts to reduce the overall testing cost, while still inheriting the exhaustive exploration of model checking to code-level testing.
We argue that this tradeoff is effective and efficient for subtle deep bugs in CRDT implementations, and hopefully in more distributed systems.

MongoDB uses Model-Based Test Case Generation (MBTCG) to ensure the equivalence between the C++ and the Golang versions of the operational transformation (OT) implementations in MongoDB Realm Sync \cite{Davis20vldb}.
\met is inspired by the MBTCG technique of MongoDB.
Moreover, the OT technique also focuses on ensuring eventual convergences among different replicas, which motivates us to first apply \met on CRDT designs and implementations.
%MBTCG is targeted at the Operational Transformation functions which ensure replicas converge to the same state.
The MBTCG technique is used at the unit testing level, and focuses on the OT module in the  Realm Sync system.
The \met framework is targeted at subtle deep bugs in CRDT data stores and the explorative testing is conducted at the system level.

%--
%--
\section{Conclusion} \label{Sec: Concl}

In this work we present the \met framework to cope with deep bugs in CRDT designs and implementations.
The developer first specifies the CRDT design in TLA+, and model checks the design to eliminate model level bugs.
Then test cases are automatically generated from the model checking traces to achieve explorative and exhaustive testing at the code level.
We demonstrate a practical application of the \met framework in the design and implementation of various data types over %RWF-RPQ and RWF-List in 
the CRDT-Redis data store. 
The \met framework not only eases the fixing of bugs which can be detected by standard testing techniques.
It also finds bugs which are out of reach of existing techniques.
We also discuss how \met increases our confidence in the correctness of CRDT designs and implementations.

In our future work, we will apply \met to more different types of distributed systems, e.g.,  distributed consensus components and atomic commit components in cloud-native distributed databases. We will also explore heuristic exploration strategies as well as parallel checking frameworks, to further tame the state explosion problem in \mett.

%--
% \end{CJK*}

%--
\begin{acks}
    This work was supported by the Cooperation Fund of Huawei-Nanjing University Next Generation Programming Innovation Lab (No. YBN2019105178SW38).
\end{acks}

%\clearpage

\balance

\bibliographystyle{ACM-Reference-Format}
\bibliography{et}

\end{document}